\documentclass[preprint,12pt]{elsarticle}

\usepackage[bookmarks=false]{hyperref}
\usepackage{amssymb}

\usepackage{bm}
\usepackage{siunitx}

\journal{International Journal of Multiphase Flow}

\begin{document}

\begin{frontmatter}



\title{Modelling large scale airgun-bubble dynamics with highly non-spherical features}

\author[1]{Shuai Li\corref{cor1}}
\author[1]{Devaraj van der Meer}
\author[2]{A-Man Zhang}
\author[1,3]{Andrea Prosperetti}
\author[1,4]{Detlef Lohse}

\cortext[cor1]{Corresponding author: s.li@utwente.nl; hgcls@163.com}

\address[1]{Physics of Fluids Group, Faculty of Science and Technology, J.M. Burgers Center for Fluid Dynamics,and MESA+ Institute, University of Twente, 7500 AE Enschede, Netherlands}
\address[2]{College of Shipbuilding Engineering, Harbin Engineering University, 145 Nantong Street, Harbin 150001, China}
\address[3]{Department of Mechanical Engineering, University of Houston, TX 77204-4006, USA}
\address[4]{Max Planck Institute for Dynamics and Self-Organization, Am Fassberg 17, 37077, G\"{o}ttingen, Germany}

\begin{abstract}

A thorough understanding of the dynamics of meter-sized airgun-bubbles is very crucial to seabed geophysical exploration. In this study, we use the boundary integral method to investigate the highly non-spherical airgun-bubble dynamics and its corresponding pressure wave emission. Moreover, a model is proposed to also consider the process of air release from the airgun port, which is found to be the most crucial factor to estimate the initial peak of the pressure wave. The numerical simulations show good agreement with experiments, in terms of non-spherical bubble shapes and pressure waves. Thereafter, the effects of the port opening time $T\rm_{open}$, airgun firing depth, heat transfer, and gravity are numerically investigated. We find that a smaller $T\rm_{open}$ leads to a more violent air release that consequently causes stronger high-frequency pressure wave emissions; however, the low-frequency pressure waves are little affected. Additionally, the non-spherical bubble dynamics is highly dependent on the Froude number $Fr$.  Starting from $Fr=2$, as $Fr$ increases, the jet contains lower kinetic energy, resulting in a stronger energy focusing of the bubble collapse itself and thus a larger pressure peak during the bubble collapse phase. For $Fr \ge 7$, the spherical bubble theory becomes an appropriate description of the airgun-bubble. The new findings of this study may provide a reference for practical operations and designing environmentally friendly airguns in the near future.

\end{abstract}


\end{frontmatter}


\section{Introduction}
\label{S:1}

To satisfy the global demand for oil, offshore oil resources become of increasing relevance \cite{Folkersen2018}. To find them airguns are widely used as a seismic source in geophysical exploration \cite{GILES1968}, as they are safe, cheap and environmentally friendly. An airgun contains a chamber filled with highly compressed air. All the ports on the chamber are opened upon firing, and the compressed air is released from the chamber into the surrounding water, thus creating a growing and then oscillating bubble. The rapid expansion and subsequent oscillations of the bubble generate pressure waves with a broad frequency spectrum. The first pressure peak has a short duration with high amplitude, which contributes more to the high-frequency waves. In recent years, zoologists found that the relatively high-frequency waves (10-150kHz) harm marine life \cite{Landr2011,Khodabandeloo2018}. This harm must be reduced and controlled at or below a safety level. The second and later pressure peaks contribute more to low-frequency waves, which can propagate far and penetrate deep into the seabed. Therefore, to meet the environmental requirement and engineering needs for the deep water seismic survey, it is necessary to design new airguns that can strengthen the energy of low-frequency waves and lessen that of high-frequency waves \cite{Chelminski2019}.

Prediction of airgun-bubble dynamics is one of the most fundamental and important problems in practical operations and designing new airguns. Many studies have been carried out in modelling the airgun-bubble dynamics with spherical bubble theories (Rayleigh-Plesset dynamics) \cite{Plesset1949,Hilgenfeldt1998JFM}. Ziolkowski \cite{Ziolkowski1970} first use the Gilmore equation to model the pressure wave emission of airgun-bubbles, followed by Johnston \cite{Johnston1982,Johnson1994}, Laws et al \cite{Laws1990}, Landr\o\ \cite{Landr1992}, Li et al \cite{LiGF2011}, de Graaf et al \cite{Graaf2014}, Zhang et al \cite{ZhangS2017}, etc. Although various physical phenomena have been incorporated into these theoretical models, there is still room for improvement. This study aims to establish an improved numerical model for airgun-bubble dynamics that differs from previous studies in two aspects as follows.

First of all, the generation and initial growth of airgun-bubbles are difficult to determine, and at the same time crucial to estimate the magnitude of the first pressure peak  \cite{Ziolkowski1970}. In some studies \cite{Ziolkowski1970,ZhangS2017}, the air release process is  ignored and the initial chamber pressure and chamber volume are taken as the initial bubble pressure and size, respectively, which inevitably over-estimates the first pressure peak. Sometimes, the initial bubble conditions are found by trial-and-error methods in case some experimental data are provided \cite{Ziolkowski1998}. Landr\o\ \cite{Landr1992} proposed a more practical model for the bubble initialization, in which the air is assumed to be ejected from the chamber into the bubble at a constant rate. de Graaf et al \cite{Graaf2014} adopted an analytical solution for the air release rate, but still overpredicted the initial pressure peak. In the present study, we solve this problem by further considering the transient port opening process, which is closer to physical reality than what had been done in previous models. This new model thus helps to find a way to predict and control the first pressure peak and high-frequency pressure waves. 

Secondly, the large-scale airgun-bubbles can hardly keep their spherical shape during their whole lifetime due to the gravity-induced pressure gradient \cite{Cox, Klaseboer2005, Zhang2015} and the Bjerknes effect of a free surface \cite{Zhang2015,Li-OE2018}. The jet formation is the main feature of a non-spherical bubble, and it has a crucial influence on the rise velocity of the bubble, the variation of the bubble volume and the far-field pressure wave \cite{Brenner2002}. Therefore, describing the large scale airgun-bubble with spherical bubble theory is found to neglect significant characteristics induced by gravity. To overcome this shortcoming, in this study, a very well verified boundary integral (BI) code \cite{Li-OE2018,Bremond2006,Bergmann2009,Bouwhuis2016} is used to study the airgun-bubble dynamics, which considers the interaction between the bubble and the ocean surface and allows for non-spherical deformation of the bubble interface. The influences of gravity, liquid compressibility and heat transfer are also incorporated in the numerical model. The boundary integral simulations agree well with experimental data and give new physical insights for airgun-bubble dynamics. 

This paper is organized as follows. First of all, we present our numerical model for airgun-bubble dynamics in Section \ref{S:2}, in which a brief discourse of the BI model and an improved scheme for the air release process are given. The validation of our model is done in Sections \ref{S:3-1}-\ref{S:3-2}, in which the non-spherical bubble motion and the pressure wave emission are compared between experiments and BI simulations. In Sections \ref{S:3-3}-\ref{S:3-6}, parametric studies are carried out to reveal the dependence of airgun-bubble dynamics on port-opening time, airgun firing depth, heat transfer and gravity. Finally, the study is summarized and conclusions are drawn in Section \ref{S:4}.

\section{Physical and numerical model}
\label{S:2}

\subsection{Airgun principle and physical problem}
\label{S:2-0}

In the field of marine geophysical exploration, different airguns have different mechanical structures \cite{Landr1992,Graaf2014,Langhammer1996}, but the working principle of general airguns can be summarized as follows: First, the chamber of a sealed airgun is filled with highly compressed air via high-pressure air hoses. During the discharging phase, an electrical signal transmitted to a solenoid valve triggers the sudden opening of the airgun port. The compressed air is then ejected from the chamber into the water, forming a growing bubble with typical diameters in the order of $O\rm(m)$. The port will close automatically once the pressure in the firing chamber decreases to a certain value, which allows the filling of air into the chamber again. Pressure waves generated by the airgun-bubble are reflected back from the interfaces that separate different stratigraphic units in the seabed. Hydrophones are used to record the amplitudes and arrival times of these waves. A comprehensive analysis of these data helps to understand the structure of the seabed \cite{Geli2009}.

\begin{figure}[htbp]
	\centering\includegraphics[width=10cm]{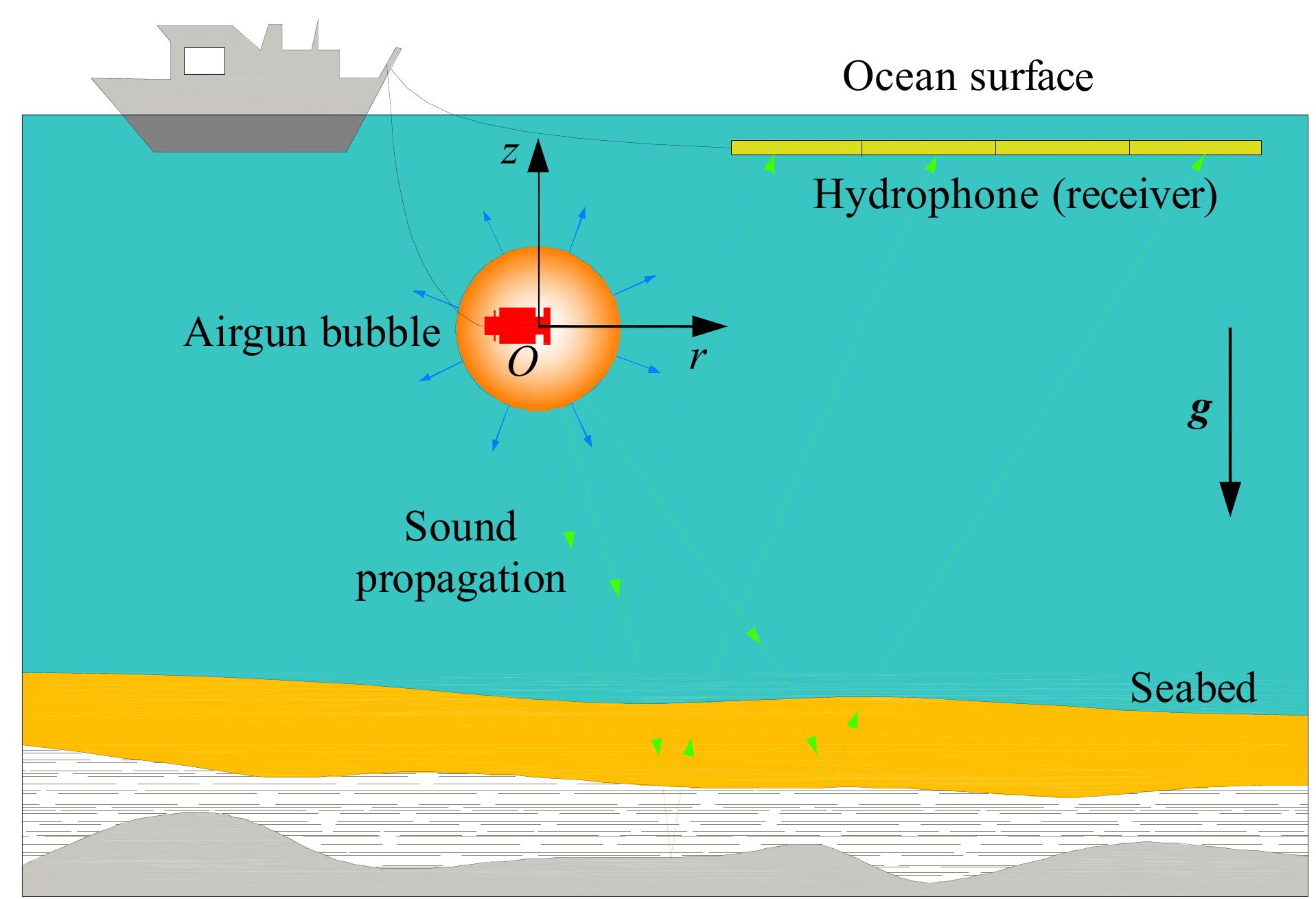}
	\caption{Principle of seismic airguns in geophysical exploration.}\label{Fig:1}
\end{figure}

In this study, we aim to model and investigate the large scale airgun-bubble dynamics with highly non-spherical features. The numerical model consists two parts, namely, the liquid flow solver (see Sections \ref*{S:2-1} and \ref*{S:2-2}) and the gas release model (see Section \ref{S:2-3}).  Both the gravity effect and the Bjerknes effect of the free surface are carefully considered, which are the main causes of non-spherical bubble motion. Note that the physical problem discussed here is restricted to an axisymmetric configuration. We define a cylindrical coordinate $(r,\theta, z)$ with the origin $O$ placed at the initial bubble (spherical shape) center and the $z$-axis pointing upward, see Figure \ref{Fig:1}. Since airguns are usually fired near the free surface and far away from the seabed, the effect of the seabed on bubble dynamics is negligible, and therefore not included in our simulations.

\subsection{Boundary integral method}
\label{S:2-1}
In the present study, the associated Reynolds number (defined as $Re=U{\rm_m}R{\rm_m}/\nu$,  where $U{\rm_m}$ is the average velocity of the bubble surface,  $R{\rm_m}$ is the maximum bubble radius, $\nu$ is the kinematic viscosity of water) can be estimated as $O(10^7)$. As such, the flow phenomenon is inertia-controlled, and viscous effects do not seem to play a role. The following airgun-bubble model is established with the potential flow theory and the boundary integral method (BI) \cite{Li-OE2018,Bremond2006,Bergmann2009,Bouwhuis2016,Oguz1993,li-jcp,wang_blake_2010}. The liquid in the near field of an airgun-bubble is assumed incompressible and inviscid, and the flow irrotational. Thus, we can describe the velocity $\textbf{\textit{u}}$ in the flow as the gradient of velocity potential $\varphi$: 
\begin{equation} 
\textbf{\textit{u}}=\nabla{\varphi},\label{Equation:1}
\end{equation}
and $\varphi$ satisfies the Laplace equation $\nabla^2{\varphi}=0$, which is equivalent to the boundary integral equation as follows:
\begin{equation} 
\alpha(\textbf{\textit{r}}){\varphi}(\textbf{\textit{r}}) = \int\left[ {\frac{\partial{\varphi}(\textbf{\textit{q}})}{\partial{n}}G(\textbf{\textit{r}},\textbf{\textit{q}})-{\varphi}(\textbf{\textit{q}})\frac{\partial{G(\textbf{\textit{r}},\textbf{\textit{q}})}}{\partial{n}}}\right] {\rm d}S{\rm_q},\label{Equation:2}
\end{equation}
where $\alpha$ is the solid angle under which the control point $\textbf{\textit{r}}$ observes the flow,   $\textbf{\textit{q}}$ is the integral point on the flow surfaces $S$, and $\textbf{\textit{n}}$ is the unit normal vector pointing out of the fluid domain. For a flat free surface case, the Green function can be taken as $G(\textbf{\textit{r}},\textbf{\textit{q}})=1/|\textbf{\textit{r}}-\textbf{\textit{q}}|-1/|\textbf{\textit{r}}-\textbf{\textit{q}}^\prime|$; for a direct simulation of the bubble-free-surface interaction, $G(\textbf{\textit{r}},\textbf{\textit{q}})=1/|\textbf{\textit{r}}-\textbf{\textit{q}}|$.
If $\varphi$ on the flow boundary (the bubble surface and the ocean surface) is known, the normal velocity $u_{n}$ can be solved afterward. The velocity tangent to the boundaries $u_\tau$ is solved by using the central-difference method. Here, the subscripts `$n$' and `$\tau$' denote the normal and tangential components, respectively. 

The bubble surface is updated by time-integrating the kinematic condition on the surface:
\begin{equation} 
\frac{{\rm d}\textbf{\textit{r}}}{{\rm d}t}=\nabla{\varphi}=u{\rm_n}\cdot\textbf{\textit{n}}+u_{\tau}\cdot\bm{\tau}.\label{Equation:3}
\end{equation}

Following Wang \& Blake\ \cite{wang_blake_2010}, we consider an acoustic correction to the traditional BI. The fluid domain is divided into two regions: the inner region near the bubble surface and the outer region. The inner region can be described by the Laplace equation (incompressible fluid) while the outer region can be described by the linear wave equation (compressible fluid). Therefore, the airgun-bubble in a weakly compressible liquid is modelled by the Laplace equation with the compressible effects appearing only in the far-field condition. More details on the mathematical derivation can be found in Wang \& Blake \ \cite{wang_blake_2010,wang_blake_2011}. Here we just give the modified dynamic boundary condition on the bubble surface:
\begin{equation} 
\frac{{\rm d}{\varphi}}{{\rm d}t}=\frac{1}{2}{|{\nabla \varphi}|^2}+\frac{p_{\infty}-p{\rm_b}}{\rho}-gz+\frac{1}{4{\pi}c}\ddot{m},\label{Equation:4}
\end{equation}
where $p{\rm_b}$ is the gas pressure on the bubble surface, $p_\infty$ is the hydrostatic pressure at $z = 0$, $g$ is the gravitational acceleration, $c$ is the sound speed, and the last term denotes the acoustic correction. Surface tension is not included in Equation (\ref{Equation:4}) since the associated Weber number can be estimated as $O(10^6)$. The quantity $m$ is defined as:
\begin{equation} 
m=\int\frac{\partial{\varphi}}{\partial{n}} {\rm d}S=\int u{\rm_n} {\rm d}S,\label{Equation:5}
\end{equation}
which is equal to the opposite value of the rate of change of the bubble volume, $m = -\dot{V}$. To avoid numerical approximation of the second time derivative of $m$, we move the last term of Equation (\ref{Equation:4}) to the left-hand side and Equation (\ref{Equation:4}) transforms into:

\begin{equation} 
\frac{{\rm d}}{{\rm d}t}\left( {\varphi}-\frac{1}{4{\pi}c}\dot{m}\right) =\frac{1}{2}{|{\nabla \varphi}|^2}+\frac{p_{\infty}-p{\rm_b}}{\rho}-gz.\label{Equation:6}
\end{equation}

The potential on the bubble surface is updated by time-integrating Equation (\ref{Equation:6}).

\subsection{Vortex ring model for toroidal bubble dynamics}
\label{S:2-2}
The jet formation is one of the most important physical phenomena for non-spherical bubbles, which is commonly seen when the bubble is subjected to strong buoyancy or becomes affected by nearby boundaries \cite{Zhang2015,Li-OE2018,Bremond2006,Blake1987,Supponen2016,Supponen2018}. The bubble becomes toroidal after the jet penetration. In this stage, the flow solution is not unique anymore and the traditional BI cannot be applied directly to simulate toroidal bubble motion. Following Wang et al \cite{Wang1996}, Zhang et al \cite{Zhang2015b} and Li et al \cite{Li-OE2018}, the latest vortex ring model is used to handle this problem. The main idea of this method is given as follows. Firstly, a vortex ring is placed inside the toroidal bubble. Its exact position is not very important as long as it is not very close to the bubble surface. The circulation of the vortex ring is set as the velocity potential-jump at the jet impact location, i.e., the potential difference between the two nodes on the axis of symmetry just before jet penetration. In the second step, the potential is decomposed into two parts: the vortex-ring induced potential $\varphi{\rm_v}$ and the single-valued remnant potential $\varphi{\rm_r}$, written as
\begin{equation} 
\varphi=\varphi{\rm_r}+\varphi{\rm_v},
\end{equation}
where the $\varphi{\rm_v}$ term can be accurately calculated from a semi-analytical method proposed by Zhang et al \cite{Zhang2015b}. Then the $\varphi{\rm_r}$ term can be obtained by subtracting $\varphi{\rm_v}$ from the total velocity potential.

The velocity is also decomposed into two parts:
\begin{equation} 
\textbf{\textit{u}}=\textbf{\textit{u}}{\rm_r}+\textbf{\textit{u}}{\rm_v}.
\end{equation}
The first part $\textbf{\textit{u}}{\rm_r}$ is the velocity caused by the remnant potential, which can be calculated from BI. The second part $\textbf{\textit{u}}{\rm_v}$ is the velocity caused by the vortex ring, which can be calculated from the Biot-Savart law.

The boundary conditions on the toroidal bubble surface are given by
\begin{equation} 
\frac{{\rm d}\textbf{\textit{r}}}{{\rm d}t}=\textbf{\textit{u}}{\rm_r}+\textbf{\textit{u}}{\rm_v},
\end{equation}

\begin{equation} 
\frac{{\rm d}}{{\rm d}t}\left( {\varphi\rm{_r}}-\frac{1}{4{\pi}c}\dot{m}\right) =\textbf{\textit{u}}\nabla{\varphi}{\rm_r}-\frac{u^2}{2}+\frac{p_{\infty}-p\rm{_b}}{\rho}-gz.
\end{equation}

\subsection{Air release model}
\label{S:2-3}

The gas pressure of the bubble interior $p_b$ is described by the ideal gas law

\begin{equation} 
p{\rm_b}=\frac{MRT}{V},\label{Equation:pb}
\end{equation}
where $M$ is the mass of the gas, $R$ is the specific gas constant, $T$ is the temperature of the gas, and $V$ is the bubble volume.

The airgun-bubble at the initial time is set as a tiny spherical bubble, with the pressure and volume set as the ambient hydrostatic pressure and a hundredth of the chamber volume, respectively. The numerical results are not sensitive to the choice of the initial bubble pressure as long as the initial bubble volume is much smaller than the chamber volume. We assume that the subsequent air flow from the airgun chamber into the bubble through the port is isentropic, the mass flow function is thus given by \cite{Graaf2014,White2003}
\begin{equation} 
\frac{{\rm d}M\rm{_b}}{{\rm d}t}=S{\rm_{port}}\sqrt{\frac{p{\rm_g}M\rm{_g}}{V\rm{_g}}\frac{2\gamma}{\gamma-1}\left[\left(\frac{p\rm{_b}}{p\rm{_g}}\right) ^{\frac{2}{\gamma}}-\left(\frac{p\rm{_b}}{p\rm{_g}}\right) ^{\frac{\gamma+1}{\gamma}}\right] },\label{Equation:dMb1}
\end{equation}
where $S{\rm_{port}}$ is the total area of the ports, $p{\rm _g}$ and $M{\rm _g}$ denote the chamber pressure and gas mass in the chamber, respectively, and $\gamma$ is a polytropic constant. 

Considering the choked flow condition, 
\begin{equation} 
\frac{p\rm{_b}}{p\rm{_g}}\leq  \left(\frac{2}{\gamma+1} \right) ^{\gamma/(\gamma-1)},
\end{equation}
the mass flow rate is bounded by 
\begin{equation} 
\frac{{\rm d}M\rm{_b}}{{\rm d}t}\leq S\rm{_{port}}\sqrt{\frac{\it p\rm_g \it M\rm_g}{\it V\rm _g}\frac{2\gamma}{\gamma-1}\left[\left(\frac{2}{\gamma+1} \right) ^{\frac{2}{\gamma-1}} -\left(\frac{2}{\gamma+1} \right) ^{\frac{\gamma+1}{\gamma-1}} \right] }.\label{Equation:dMb2}
\end{equation}

During the air-release stage, the bubble is an open thermodynamical system and we assume this process is quasi-static. Hence, the temperature variation of the bubble gas can be derived from the first law of thermodynamics:
\begin{equation} 
\frac{{\rm d}{T\rm{_b}}}{{\rm d}t}=\frac{1}{M{\rm_b}c\rm{_v}}\left(c{\rm_p}T{\rm_g}\frac{{\rm d}{M\rm{_b}}}{{\rm d}t}-c{\rm_v}T{\rm_b}\frac{{\rm d}{M\rm{_b}}}{{\rm d}t}-\frac{{\rm d}{Q}}{{\rm d}t}-p{\rm_b}\frac{{\rm d}{V\rm{_b}}}{{\rm d}t}\right), \label{Equation:15}
\end{equation}
where $T{\rm_b}$ and $T{\rm_g}$ denote the temperature in the bubble and chamber, respectively, $c\rm{_v}$ and $c\rm{_p}$ are the specific heat capacity of the gas at constant volume and pressure, respectively, and the $\frac{{\rm d}{Q}}{{\rm d}t}$ term denotes the rate of heat conduction across the bubble surface, which is written as
\begin{equation} 
\frac{{\rm d}{Q}}{{\rm d}t}={\kappa}(T{\rm_b}-T{\rm_w})S{\rm_b},\label{Equation:16}
\end{equation}
where $\kappa$ is the heat transfer coefficient, and $T{\rm_w}$ is the temperature of surrounding water. $\kappa$ is assumed to be a constant in this study. As suggested in the literatures\cite{Graaf2014,ni2011}, the value of $\kappa$ in Equation (\ref{Equation:16}) ranges from 2000 to 8000 $\rm {W/m^2K}$ for conventional airgun-bubbles. The effects of this term will be discussed in Section \ref{S:3-5}.

The temperature in the chamber $T{\rm_g}$ can be updated using the energy conservation law. The total energy of the whole system keeps a constant, given by 
\begin{equation} 
E_0=M{\rm_{g0}}c{\rm_v}T{\rm_{g0}}+M{\rm_{b0}}c{\rm_v}T{\rm_{b0}},
\end{equation}
where the subscript ``0'' denotes the initial value.

At each time step, the energy associated with the bubble can be calculated from
\begin{equation} 
E{\rm_b}=M{\rm_{b}}c{\rm_v}T{\rm_b}+Q+p_{\infty}V{\rm_b}+E{\rm_k}+E{\rm_{acoustic}},\label{Equation:Eb}
\end{equation}
where the five terms on the right-side denote the internal energy of the bubble gas, the heat transferred across the bubble surface into water, the potential energy, the kinetic energy of water and the acoustic radiation energy, respectively. The first three terms can be easily calculated using the current physical quantities. The last two terms are given by \cite{WangQX2016JFM}
\begin{equation} 
E{\rm_k}=\frac{1}{2} \rho\int_S{\varphi\frac{{\partial}{\varphi}}{{\partial}n}}{\rm d}S,\label{Equation:Ek}
\end{equation}
\begin{equation} 
E{\rm_{acoustic}}=\frac{\rho}{4\pi c}\left[\dot{V}(0)\ddot{V}(0)-\dot{V}(t)\ddot{V}(t)+   \int_0^t{\ddot{V}^2(t)}{\rm d}S \right]. \label{Equation:acoustic}
\end{equation}

The energy of the gas in the chamber can be easily calculated by subtracting $E{\rm_b}$ from $E_0$ and the temperature in the chamber is written as
\begin{equation} 
T{\rm{_g}}=\frac{E_0-E{\rm_b}}{M{\rm_{g}}c{\rm_v}}.\label{Equation:Tg}
\end{equation}

Since the gas-release phase is within a short time, the heat transfer effect between the gas in the chamber and the airgun body is neglected.

\subsection{Computation of the pressure wave}
\label{S:2-4}
The dynamic pressure in the near field of the bubble can be obtained from the unsteady Bernoulli equation:
\begin{equation} 
p{\rm_d}=-\rho\left(\frac{\partial{\varphi}}{\partial{t}}+\frac{u^2}{2} \right),\label{Equation:pd}
\end{equation}

For the far-field, the pressure wave generated by the oscillating bubble is calculated as \cite{Brennen2015}
\begin{equation} 
p{\rm_s}=\frac{\rho}{4{\pi}D}\frac{{\rm d^2}{V}}{{\rm d}t^2}.\label{Equation:ps}
\end{equation}
where $D$ is the distance between the pressure measurement point and the bubble center. It is easy to see that the above two equations are equivalent for a far-field point since the velocity term in Equation (\ref{Equation:pd}) is proportional to $1/D^2$ and negligible when compared with the first term. In the present study, Equation (\ref{Equation:pd}) is used to calculate the pressure in the near field of the bubble and Equation (\ref{Equation:ps}) is used to calculate the pressure wave in the far-field. Note that Equation (\ref{Equation:ps}) does not account for any variation of the pressure in the azimuthal direction. Supponen et al \cite{Supponen2019prf} found that some shock waves emitted at the collapse of laser-induced cavitation bubbles, especially the jet impact shock, might have evidence of some directionality in the near field. However, their experiments also suggest that this directionality must be subdominant in the far-field. In the present study, the bubble jet impact velocity is around 30 m/s, which is much smaller than that in Supponen et al \cite{Supponen2019prf}. Therefore, the directionality of the pressure wave generated by the airgun-bubble is expected to be small and using Equation (\ref{Equation:ps}) to calculate the far-field pressure should be appropriate.

de Graaf et al \cite{Graaf2014} found that the initial pressure peak can easily be over-estimated if the real throttling (port) area is used in calculating the air release rate. Therefore these authors used a reduced area in their model. In the present study, we argue that this problem can be solved by considering the port opening process, which is physically a more realistic approach than that of previous models. Here we simply assume the port area to increase linearly within a short time $T_{\rm open}$, i.e.,
\begin{equation} 
S_{\rm port}=\left\{
\begin{array}{rcl}
	S_{\rm max}\cdot t/T_{\rm open}       &      & {\ \ \ \ \ 0< t      <      T_{\rm open}}\\
	S_{\rm max}     &      & {T_{\rm open} \leq t < T_{\rm close}}\\
	0     &      & {T_{\rm close} \leq t < \infty}\\
\end{array}, \right. \label{Equation:Sport}
\end{equation}
where $S_{\rm max}$ is the maximum area of the port, and $T_{\rm close}$ is the time at which the port closes. In this study, the port closure is triggered when 95 \% of the air in the chamber is released. Here we don't consider the port close process because the air release rate is very small at the final air release stage.

\subsection{Flowchart of the simulation}
\label{S:2-6}
The foregoing sections have given the models for the water and gas separately. This section gives the numerical procedure for calculation of the airgun-bubble dynamics:
\begin{enumerate}[\indent(1)]	
	\item Read input data and initialization;
	\item	Begin time stepping;
	\item	Calculate the velocity on the bubble surface $\textbf{\textit{u}}$ using BI; 
	\item	Calculate the port area $S_{\rm port}$ using Equation (\ref{Equation:Sport}) and get the air flow rate from Equations (\ref{Equation:dMb1}-\ref{Equation:dMb2});
	\item	Update the bubble mass and calculate the gas pressure inside the bubble using Equation (\ref{Equation:pb});
	\item	Use the dynamic boundary condition (Equation \ref{Equation:6}) to update the potential on the bubble surface;
	\item	Use the kinematic boundary condition (Equation \ref{Equation:3}) to update the position of the bubble surface;
	\item	Update the temperature of the bubble gas using Equation (\ref{Equation:15});
	\item	Solve energy equations (\ref{Equation:Eb}-\ref{Equation:acoustic}) and get the temperature of the gas in the chamber using Equation (\ref{Equation:Tg});
	\item	Obtain the pressure field using Equation (\ref{Equation:pd}) or (\ref{Equation:ps});
	\item	Increment time and go back to Step 2.
\end{enumerate}
During the simulation, the adaptive time step is determined according to Wang et al \cite{Wang1996} and Zhang et al \cite{ZhangS2017}. To improve accuracy, the second-order Runge-Kutta method is adopted for the forward time integration. In addition, the five-point smoothing technique and spline interpolation \cite{Pearson2004} are used to maintain the stability of the simulation.

\section{Results and discussions}
\label{S:3}

\subsection{Comparison of the non-spherical bubble motion between experimental observation and BI simulation}
\label{S:3-1}
To validate the numerical model in simulating non-spherical bubbles, we compare our numerical results with those from an experiment carried out for an electric discharge bubble in a low-pressure tank, captured by a high-speed camera working at 15 000 frames per second. For more details on the experimental setup, we refer to Zhang et al \cite{Zhang2015}. Here we only give the parameters used in that experiment: the air pressure in the sealed tank is reduced to 2000 Pa and the bubble is generated at the water depth of 260 mm. The maximum bubble radius $R\rm_m$ reaches 50 mm. The spatial resolution of the experimental images is 0.37 mm per pixel. 

\begin{figure}[htbp]
	\centering\includegraphics[width=13cm]{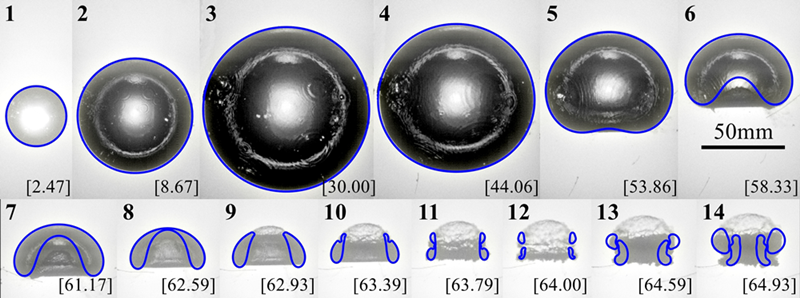}
	\caption{Comparison of the non-spherical bubble shapes (side view) between experiment \cite{Zhang2015} and boundary integral simulation (denoted by the blue lines). In the experiment, the bubble is generated by the underwater electric discharge at the water depth of 260 mm in a closed tank with the air pressure reduced to 2000 Pa. The associated Froude number (defined as $Fr = \sqrt{p_\infty/{\rho gR\rm_m}}$) reaches 2.12. The bubble keeps a spherical shape during the first expansion (frames 1-3) and a liquid jet forms during the collapse phase (frames 4-8). The bubble transforms into a toroidal form after the jet penetration (frames 9-10) and the splitting of the toroidal bubble is observed at the final collapse phase of the bubble (frames 11-12), followed by the rebounding phase of bubbles (frames 13-14). The times of each frame are given in the square brackets	(unit: ms).}\label{Fig:2}
\end{figure}

In the numerical simulation, the bubble is assumed to initiate immediately with high pressure gas inside. As previously done by Tong et al \cite{Blake1999}, Klaseboer et al \cite{Klaseboer2006}, Goh et al \cite{Goh2017}, Hsiao et al \cite{Hsiao2013} and Koukouvinis et al \cite{Koukouvinis2016}, the initial pressure of the bubble $p{\rm_{b0}}$ is set as 500 times the ambient hydrostatic pressure and the initial bubble radius is set as $0.084R_m$, which is estimated using the following equation \cite{Plesset1949,Klaseboer2005}. 
\begin{equation} 
\frac{p\rm{_{b0}}}{p_{\infty}}\left[\left(\frac{R_0}{R\rm_m} \right) ^{3\gamma}-\left(\frac{R_0}{R\rm_m} \right) ^{3} \right]  = (\gamma-1)\left[\left(\frac{R_0}{R\rm_m} \right) ^{3}-1\right].
\end{equation}
It has been demonstrated that the bubble motion is not sensitive to the choice of the initial pressure (the initial bubble radius changes accordingly for reaching a specified maximum bubble radius) \cite{Li-OE2018,Blake1999}. Since the associated Mach number in this experiment is below 0.02, the compressibility of the liquid can be neglected. The heat transfer is also ignored in this case and the polytropic constant is taken as 1.25 \cite{Li-OE2018,Goh2017}. 

Figure \ref{Fig:2} shows the bubble motion during the first cycle and the early second cycle, in which each experimental image (side view) is overlaid with the numerical results (denoted by the blue lines). As can be seen, the bubble expands spherically (frames 1-3) and the bubble bottom becomes flattened during the early collapse phase due to the pressure gradient caused by gravity (frames 4-5). Thereafter, a pronounced liquid jet forms from the bubble bottom (frames 6-7) and impacts on the upper surface of the bubble (frame 8). Since the curved bubble surface acts as a divergent lens, the liquid jet looks smaller by a factor of 0.75 \cite{Li-OE2018,Philipp1998}, which is mainly responsible for the discrepancy of jet profiles between the experimental observations and the numerical simulation. After the jet penetration (frame 9), the flow domain transforms from single-connected to double-connected. The bubble becomes a toroidal bubble and continues shrinking afterward. It is worth noting that an annular jet appears on the outer surface of the bubble (frame 10) and propagates downward, which finally leads to the bubble split (frame 11). The multiple-vortex-ring model \cite{Zhang2015b} is used to simulate the subsequent interaction between toroidal bubbles. The minimum volume of the bubble is reached around frame 12, followed by the rebounding of the bubbles (frames 13-14). It is observed that the upper bubble expands faster in the radial direction and the lower bubble is sucked in by the upper bubble. This phenomenon is similar to the leapfrogging of vortex rings \cite{Cheng2015}, however, the two toroidal bubbles here presumably merge into a bigger bubble during the rebounding phase. A good agreement is achieved between the BI simulation and the experimental observation, including the bubble expansion, collapse, jetting, toroidal bubble splitting, and the interaction between two toroidal bubbles.

\subsection{Comparison of the pressure wave generated by an airgun-bubble between experiments and BI simulation}
\label{S:3-2}
In this section, we aim to simulate real airgun-bubbles (SERCEL type 520 airgun) by using the BI code. The experiments we use as reference were conducted by the Australian Defence Science and Technology Organisation and the pressure wave generated by a single airgun-bubble was measured \cite{Graaf2014}. Two experiments were reported, in which the airgun was fired at different air pressures (20.7 MPa and 17.2 MPa, respectively) and different airgun firing depths (3 m and 5m, respectively). Other parameters were kept the same, and given as follows: the chamber volume was 8521 cm$^3$, the port area was 128 cm$^2$, and the pressure gauge was placed at the same depth with the airgun at a position of 2.22 m away from the airgun center. As far as we are concerned, there is no report on the value of the opening time of the airgun valve $T_{\rm open}$ in the literature. We find that satisfactory and reasonable results can be achieved if $T_{\rm open}$ is set 4 ms for the present type of airgun. More discussion on the effect of $T_{\rm open}$ will be given in Section \ref{S:3-3}. As suggested by de Graaf et al \cite{Graaf2014}, the heat transfer plays a role in airgun-bubble dynamics, which might be enhanced by the increase of bubble surface area due to the development of the Rayleigh-Taylor instability on the bubble surface \cite{Graaf2014b}. The heat transfer coefficient ${\kappa}$ in Equation (\ref{Equation:16}) is set as 7000 W/m$^2$K \cite{Graaf2014}. In this and the subsequent simulations of airgun-bubbles, the polytropic constant $\gamma$ is set as 1.4.

\begin{figure}[htbp]
	\centering	
	\includegraphics[width=12cm]{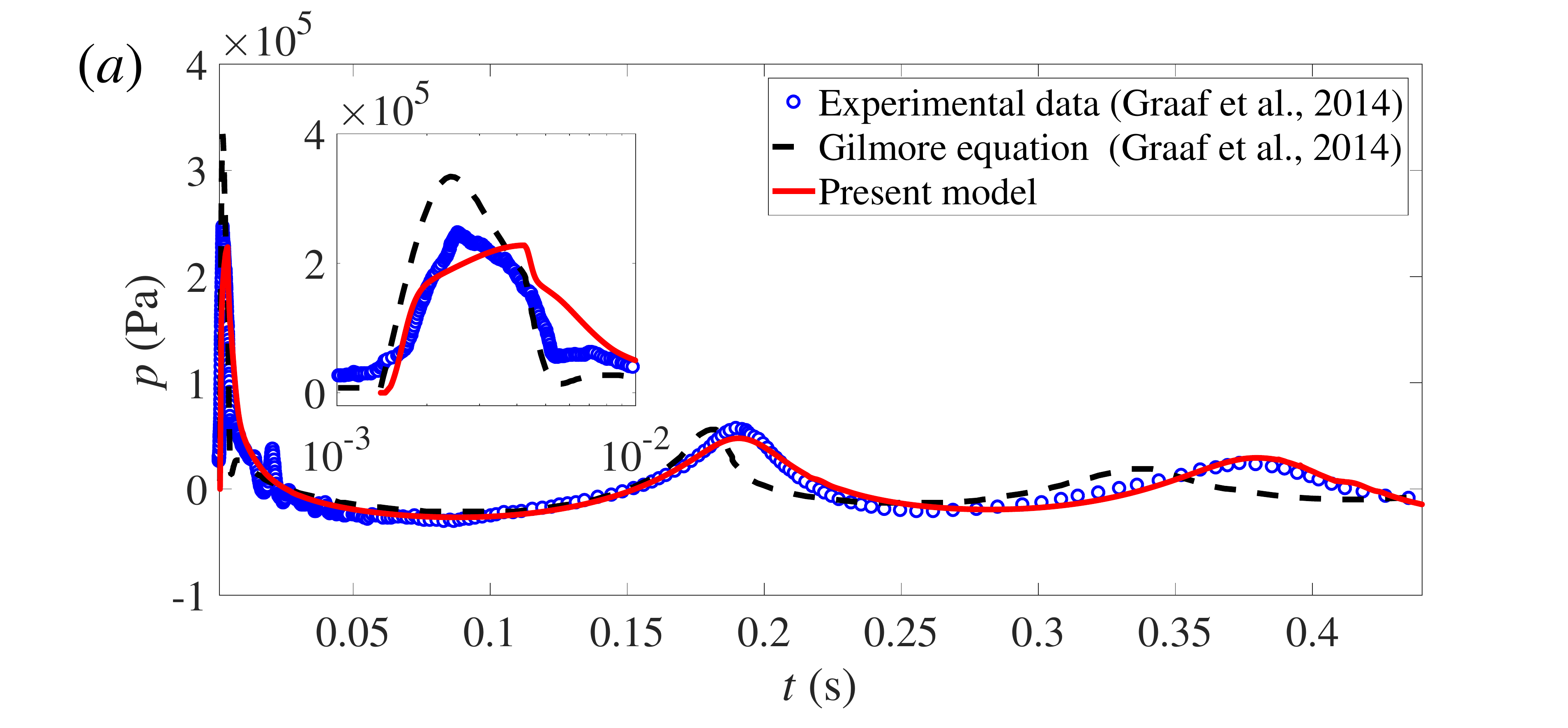}
	\includegraphics[width=12cm]{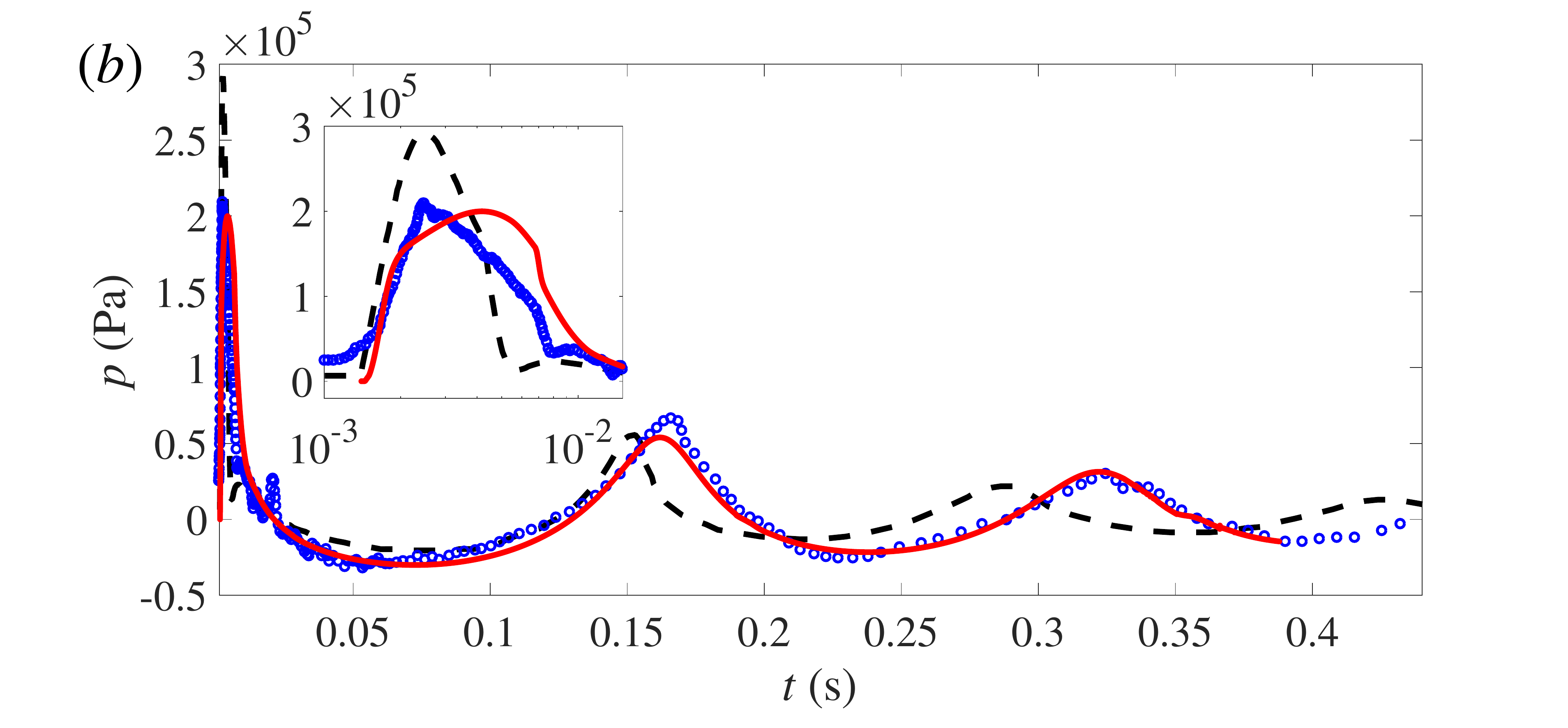}
	\caption{Comparison of the pressure waves between the theoretical model based on the Gilmore equation \cite{Graaf2014} (denoted by black dashed lines), the present numerical model (denoted by red solid lines) and experimental measurements (denoted by blue circles). (a) The pressure of the airgun chamber is 20.7 MPa and the airgun firing depth is 3 m, (b) The pressure of the airgun chamber is 17.2 MPa and the airgun firing depth is 5 m. In both cases, the sensor was installed at the same depth as the airgun center and 2.22m away. In the insets, the data are plotted using a logarithmic time scale to highlight the initial pressure wave. }\label{Fig:3}
\end{figure}

Figure \ref{Fig:3}(a) shows the comparison of the pressure wave between the first experiment (denoted by blue circles) and BI simulation (denoted by the red solid line), and the theoretical result obtained by de Graaf et al \cite{Graaf2014} is also given (denoted by the black dashed line). The results shown here have been superposed with the  “ghost signal”, i.e., the reflection of the pressure wave from the free surface. Here, the reflection coefficient is taken as -1 \cite{Graaf2014,Cox}. Clearly, the first pressure peak is well predicted by the present model but over-estimated by the theoretical model without considering the port opening process. The first pressure peak is reached within 3 ms, which is at a very early stage of the air release process.  There exists an evident sharp reduction of the pressure curve around 4 ms, which is attributed to the “ghost signal”. In the experimental signal, the pressure reflection from the tank wall is responsible for the bump at $t$ = 21 ms. It is noted that the bubble period and the subsequent two pressure peaks are well predicted by the present model. However, there exists an obvious difference between the theoretical model and the experiment, which is attributed to the neglect of the non-spherical features of bubble motion in spherical bubble theory. As the airgun volume increases, the bubble will lose its spherical shape earlier during the collapse phase, and spherical bubble theories are expected to deviate more from reality. The stronger gravity effect (buoyancy effect) on the airgun-bubble deserves further investigation and will be discussed in Section \ref{S:3-6}. 

The second experimental case with different air pressure and airgun firing depth is calculated using the same numerical setup, as shown in Figure \ref{Fig:3}(b). The first pressure peak is 39\% over-estimated by the theoretical model and 4.3\% under-estimated by the present model. Some uncertainties in practical operations or measuring error might be responsible for the slight difference. For example, the port opening time is not accurately fixed every time, which might depend on the air pressure and needs careful measurement in practical operation. Still, the bubble period and the subsequent two pressure peaks are well reproduced by the BI model. Apparently, the theoretical prediction is very different from the experimental result, indicating the distinct advantage of the present numerical model over the spherical bubble theory.

\begin{figure}[htbp]
	\centering	
	\includegraphics[width=4.2cm]{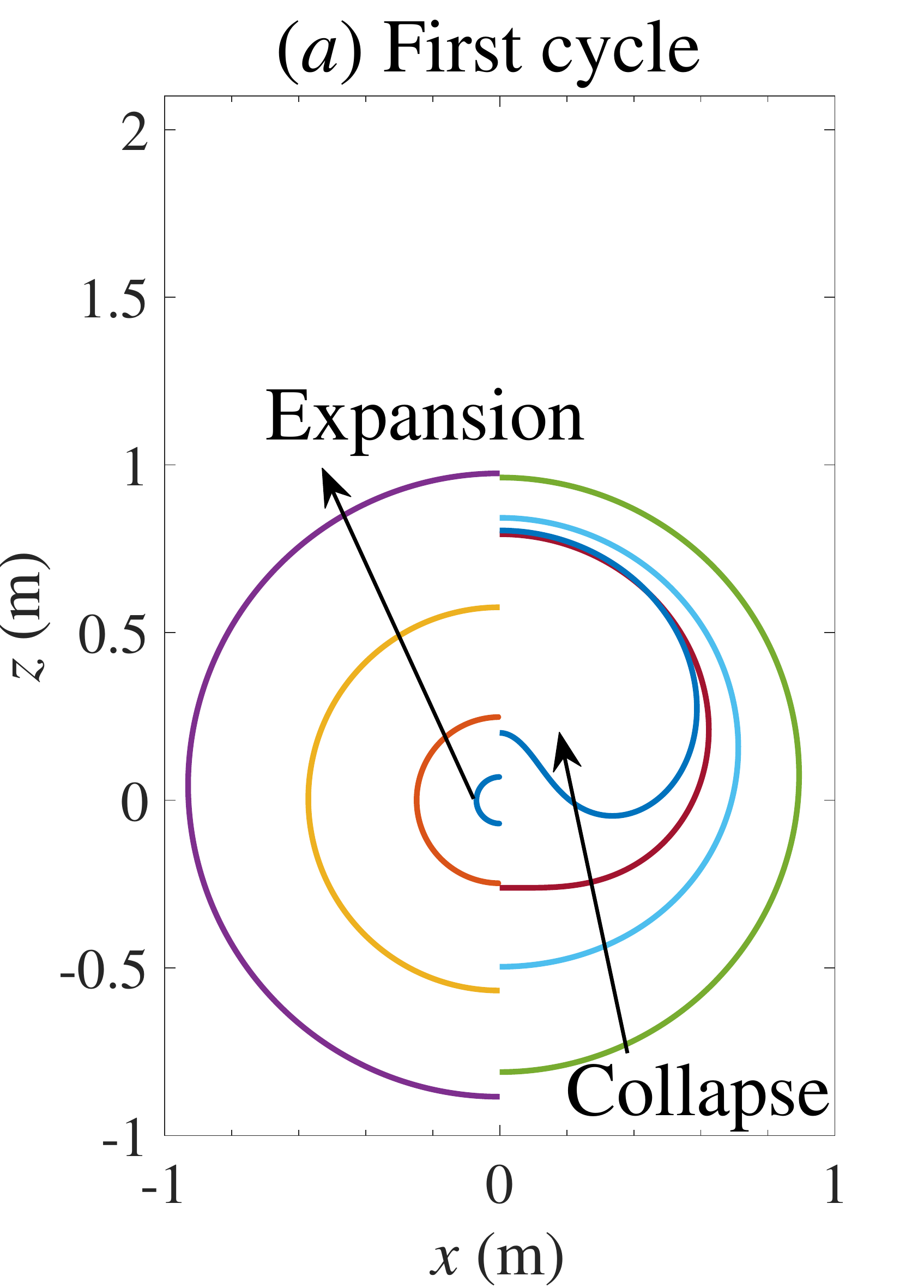}
	\includegraphics[width=4.2cm]{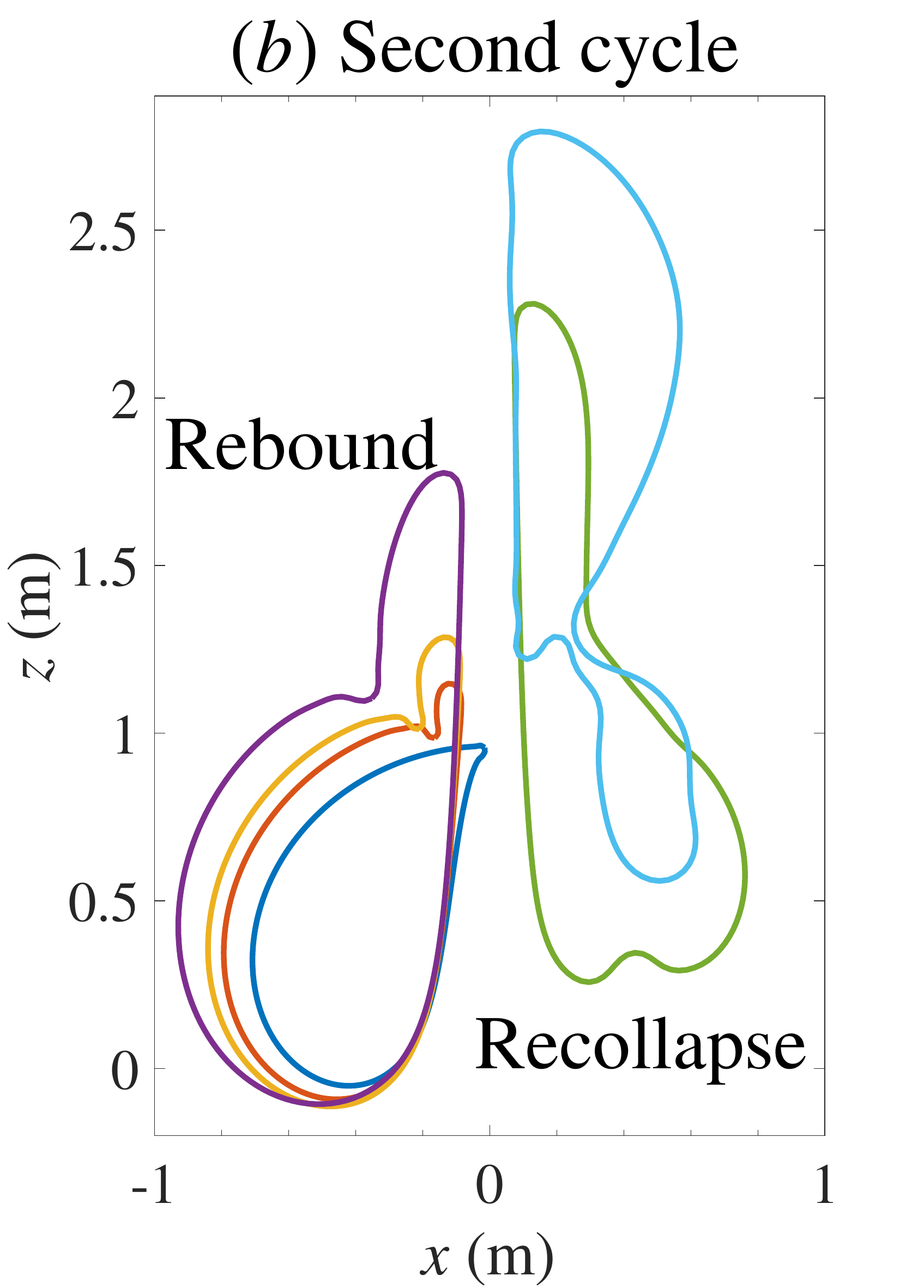}
	\includegraphics[width=4.2cm]{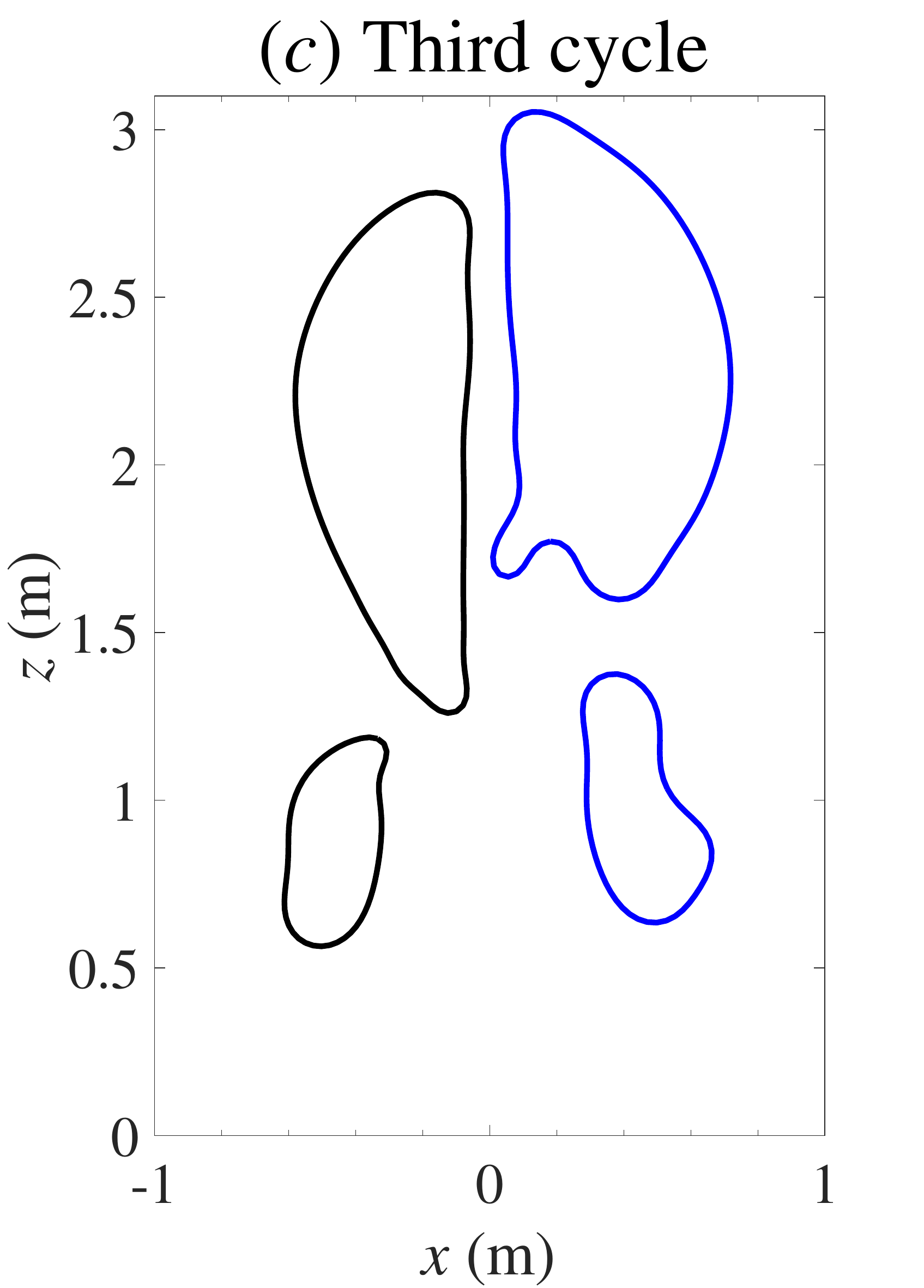}
	\caption{Evolution of the non-spherical bubble shape for the same case as in Figure \ref{Fig:3}(a). Panels (a)-(c) give the bubble motion during the first, second and third cycles of the bubble, respectively. The frame times for panel (a) are 0.001, 0.005, 0.020, 0.092, 0.120, 0.160, 0.175, 0.192, and the ones for panel (b) are 0.215, 0.230, 0.240, 0.284, 0.35, 0.408 and the ones for panel (c) are 0.410 and 0.440 (unit: second).}\label{Fig:4}
\end{figure}

For a variety of reasons, there are few reports on the high-speed photography of airgun-bubbles \cite{Graaf2014b,Langhammer1996}. Thus the dynamic behavior of airgun-bubbles is still not clear. Figure \ref{Fig:4} gives the evolution of the bubble shapes for the first experimental case. The bubble keeps a spherical shape during most of the first oscillation cycle and the non-spherical features start to develop at the final stage of the bubble collapse phase, as shown in Figure \ref{Fig:4}(a). The jet penetration occurs at the rebounding stage and the toroidal bubble is thus created and elongated in the vertical direction, as shown in the left half of Figure \ref{Fig:4}(b). During the recollapse phase of the bubble, a second liquid jet also forms from the bubble bottom and impacts on the side surface of the toroidal bubble, as shown in the right half of Figure \ref{Fig:4}(b). The minimum value of the sphericity \cite{Wadell1935} of the bubble (defined as $\pi^{1/3}(6V)^{2/3}/A$, where $A$ is the surface area of the bubble) is below 0.6, indicating highly non-spherical features of the bubble. The subsequent interaction between two sub-toroidal bubbles is simulated by the multiple-vortex-ring model \cite{Zhang2015b}. The total volume of bubbles varies smoothly after the splitting, thus there is no evident pressure jump in the flow field around the splitting moment ($t=0.408\rm s$), as shown in Figure \ref{Fig:3}. The splitting of the bubble might be a mechanism that determines the bubble mass loss and energy dissipation of airgun-bubbles \cite{Langhammer1996}. 

\subsection{Sensitivity study of the port opening time}
\label{S:3-3}
\begin{figure}[htbp]
	\centering	
	\includegraphics[width=6.5cm]{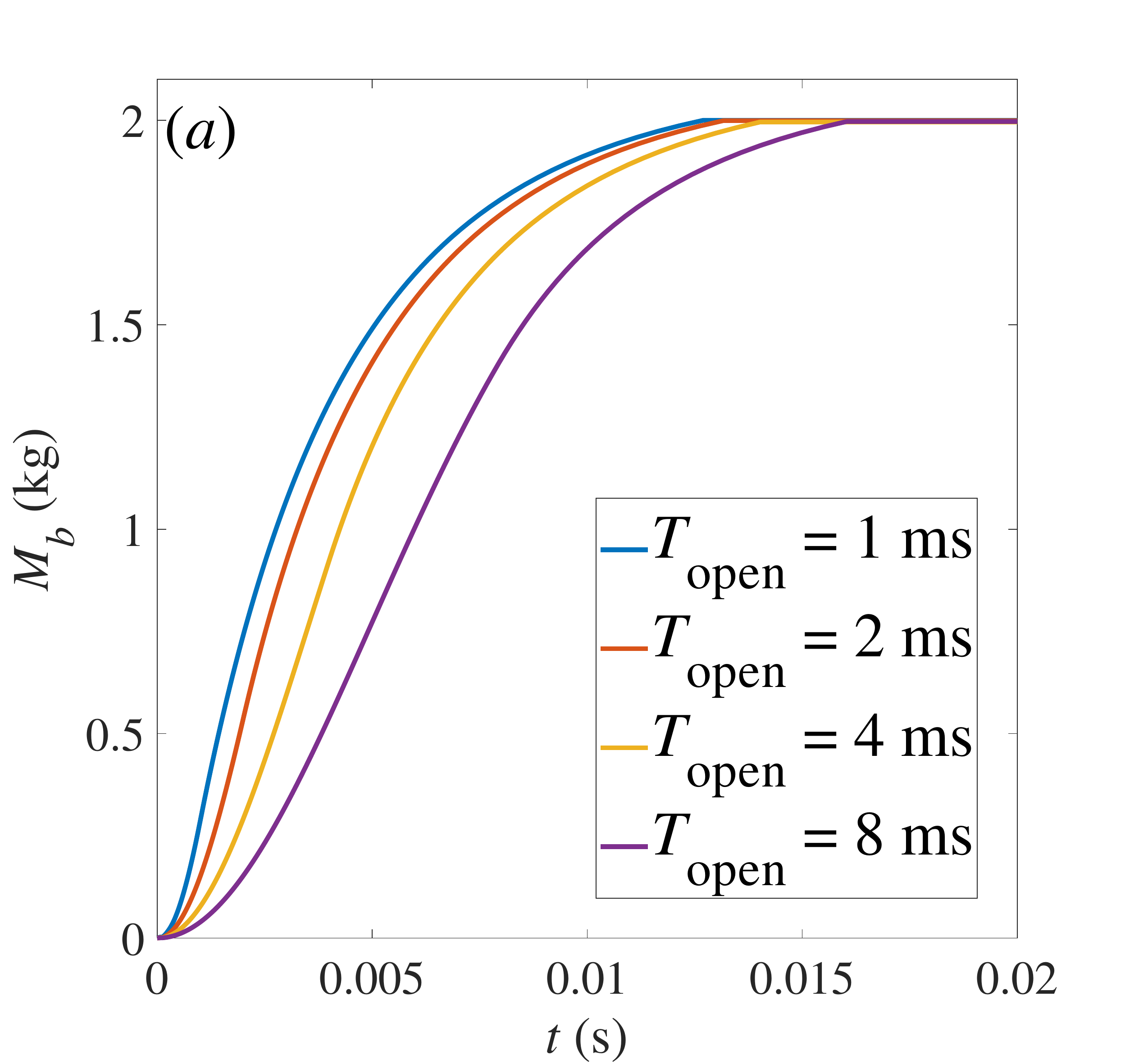}
	\includegraphics[width=6.5cm]{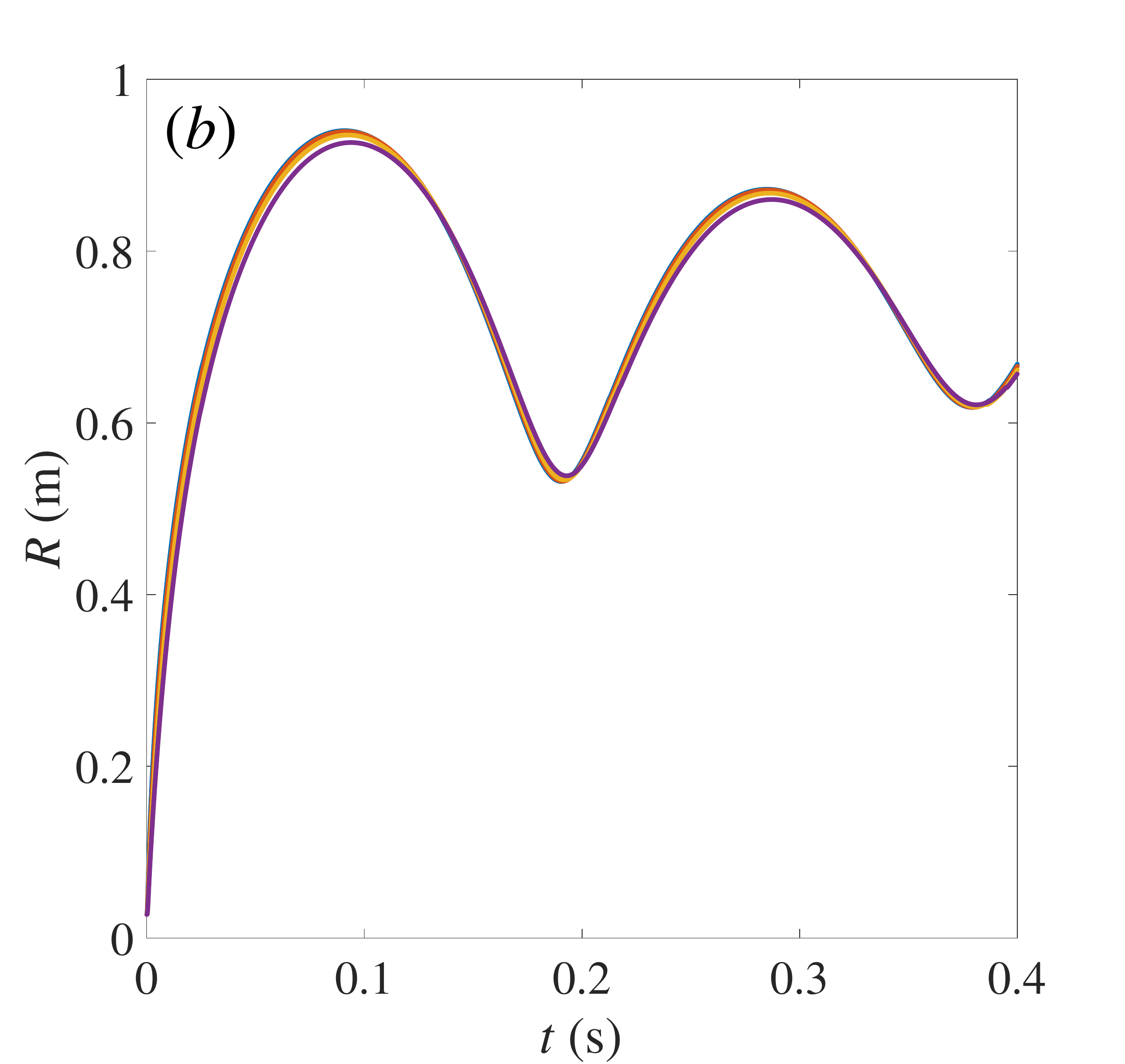}
	\includegraphics[width=6.5cm]{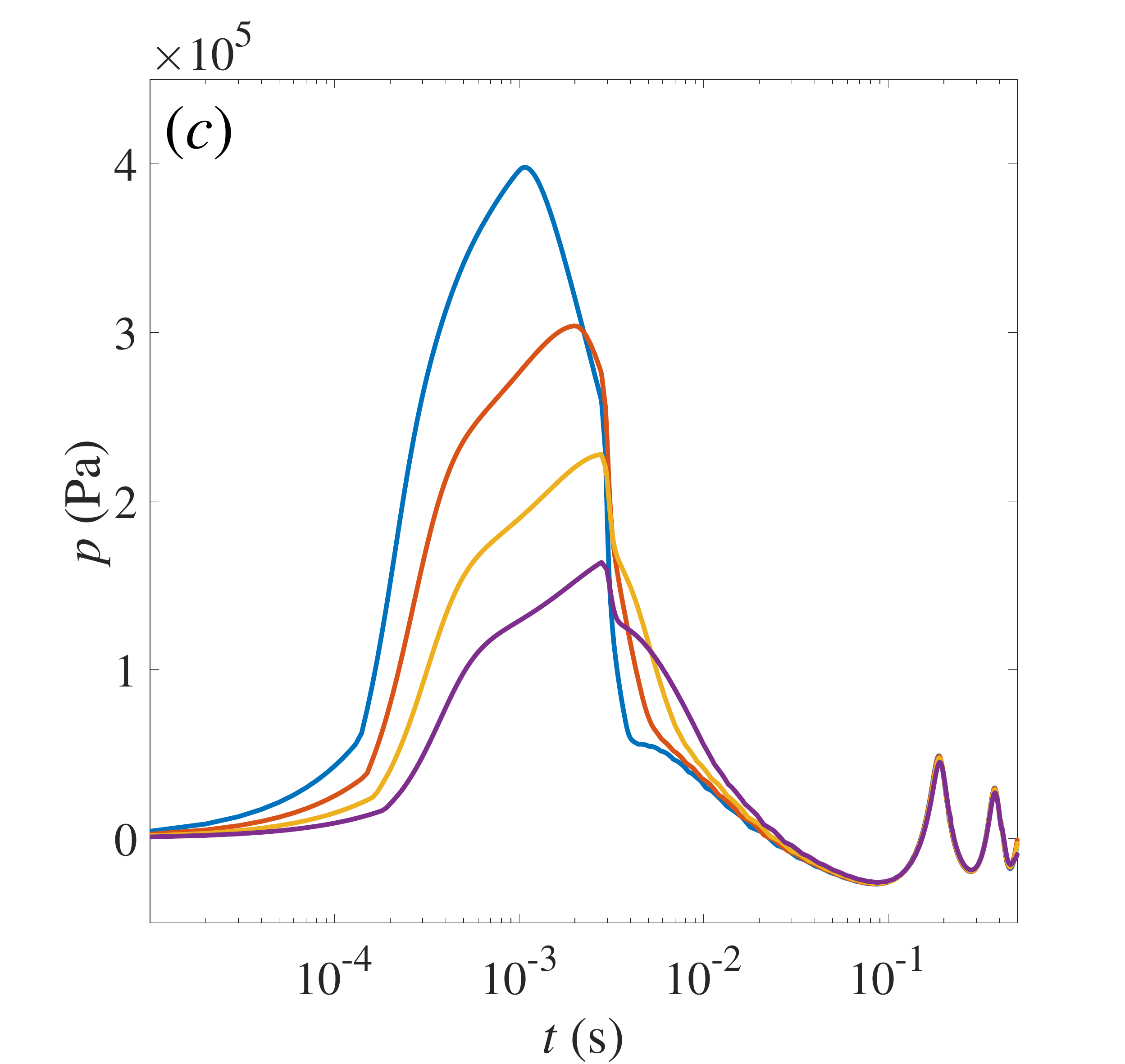}
	\includegraphics[width=6.5cm]{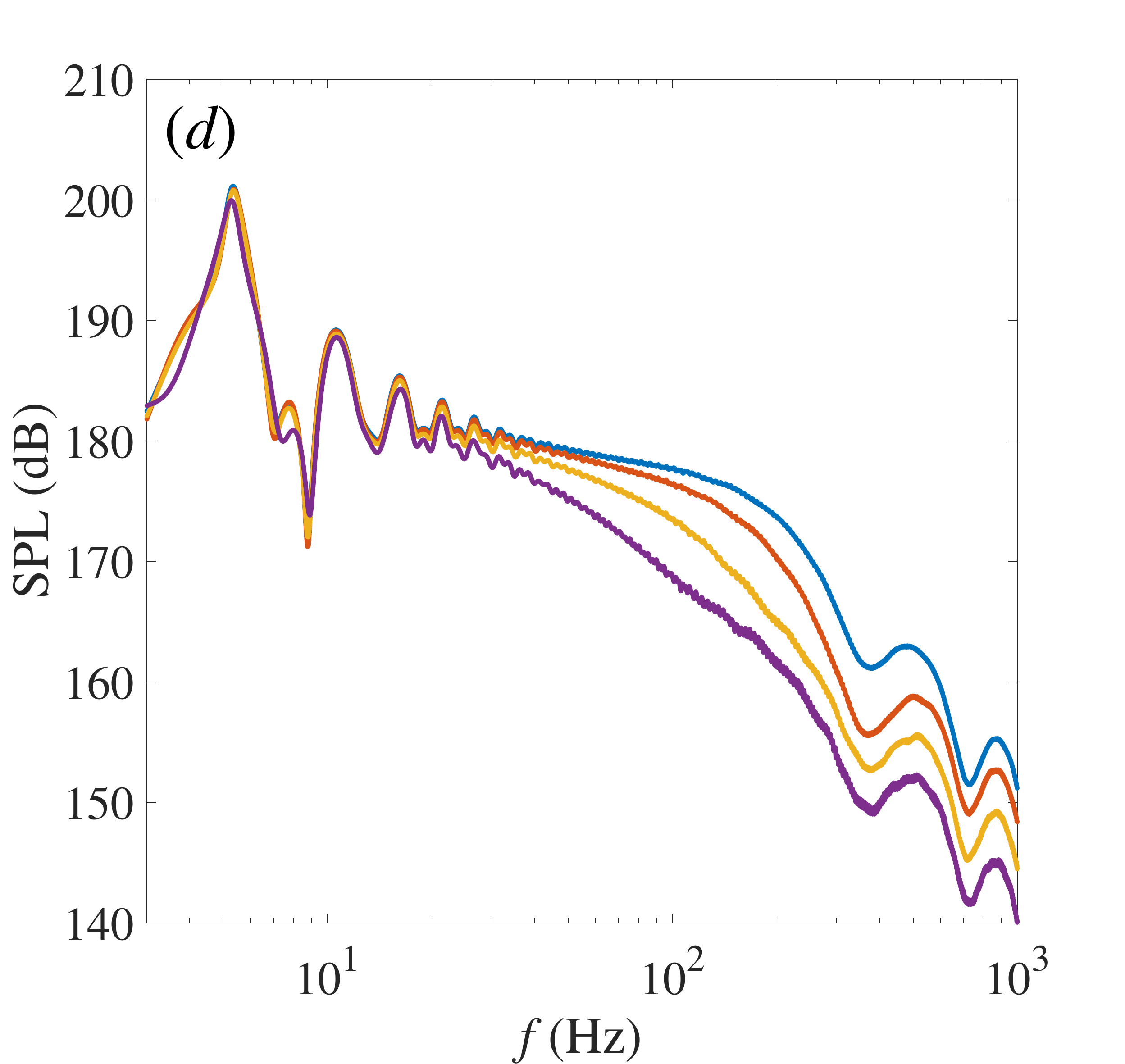}
	\caption{The effects of the port opening time. (a) Bubble mass, (b) equivalent bubble radius, (c) pressure wave at the same measure point as that in Figure \ref{Fig:3}(a), (d) pressure spectrum. The parameters are the same to the case in Figure \ref{Fig:3}(a).}\label{Fig:5}
\end{figure}

As stated in Section \ref{S:3-2}, the time $T_{\rm open}$ in which the port of airgun fully opens is an important factor that controls the initial air release rate, which consequently affects the first pressure peak. There is little knowledge on the effect of $T_{\rm open}$ in previously published literature. Thus a sensitivity study of $T_{\rm open}$ will be conducted in this section based on the first experimental case (referred to as the ``standard case'' in this and subsequent sections). In this study, we simply assume the port area to increase linearly within $T_{\rm open}$. Four different simulations are carried out with $T_{\rm open}$ being 1, 2, 4 and 8 ms, respectively. Other parameters are kept the same as that in Figure \ref{Fig:3}(a). Figure \ref{Fig:5} shows the comparisons of the numerical results between these cases. As shown in Figure \ref{Fig:5}(a), the air mass in the bubble increases faster with a smaller $T_{\rm open}$ and the total bubble mass gets higher within the same air release time. As a result, the bubble achieves a larger radius (see Figure \ref{Fig:5}(b)) and the maximum difference between these four cases is around 2\%. The differences in the bubble period and the second pressure peak are within 1\% and 8\%, respectively. However, the difference in the first pressure peak is over 140\%, as shown in Figure \ref{Fig:5}(c). That is to say,  the port opening time mainly and significantly affects the first pressure peak. The sound pressure levels (defined as SPL = 20log$(pD/p_0D_0)$, where the reference pressure $p_0$ is taken as 1 $\rm\mu$Pa and the reference distance $D_0$ = 1 m) from the pressure waves in Figure \ref{Fig:5}(c) are given in Figure \ref{Fig:5}(d). The magnitude of the SPL in the low-frequency range ($f < 20 \rm Hz$) does not change much with varying $T_{\rm open}$; however, a significant difference can be observed as the frequency increases. This implies that a faster port opening process contributes more to higher frequency pressure waves but the effect of $T_{\rm open}$ on low-frequency pressure waves is insignificant. We also found that the first peak on the SPL curve is reached when $f = 5.3\ \rm Hz$, which is inconsistent with the resonant frequency of the bubble \cite{Brenner2002,Brennen1995}, given by
\begin{equation} 
f=\frac{1}{2\pi R_{\rm e}}\sqrt{\frac{3\gamma p_{\infty}}{\rho}},\label{Equation:21}
\end{equation}
where $R_{\rm e}$ is the equilibrium bubble radius at the ambient pressure $p_{\infty}$. From the discussion above, we can draw the conclusion that the port opening time $T_{\rm open}$ is a key to control the emission of high-frequency pressure waves. This provides references for the future design of environmentally friendly airguns.

\begin{figure}[htbp]
	\centering	
	\includegraphics[width=6.5cm]{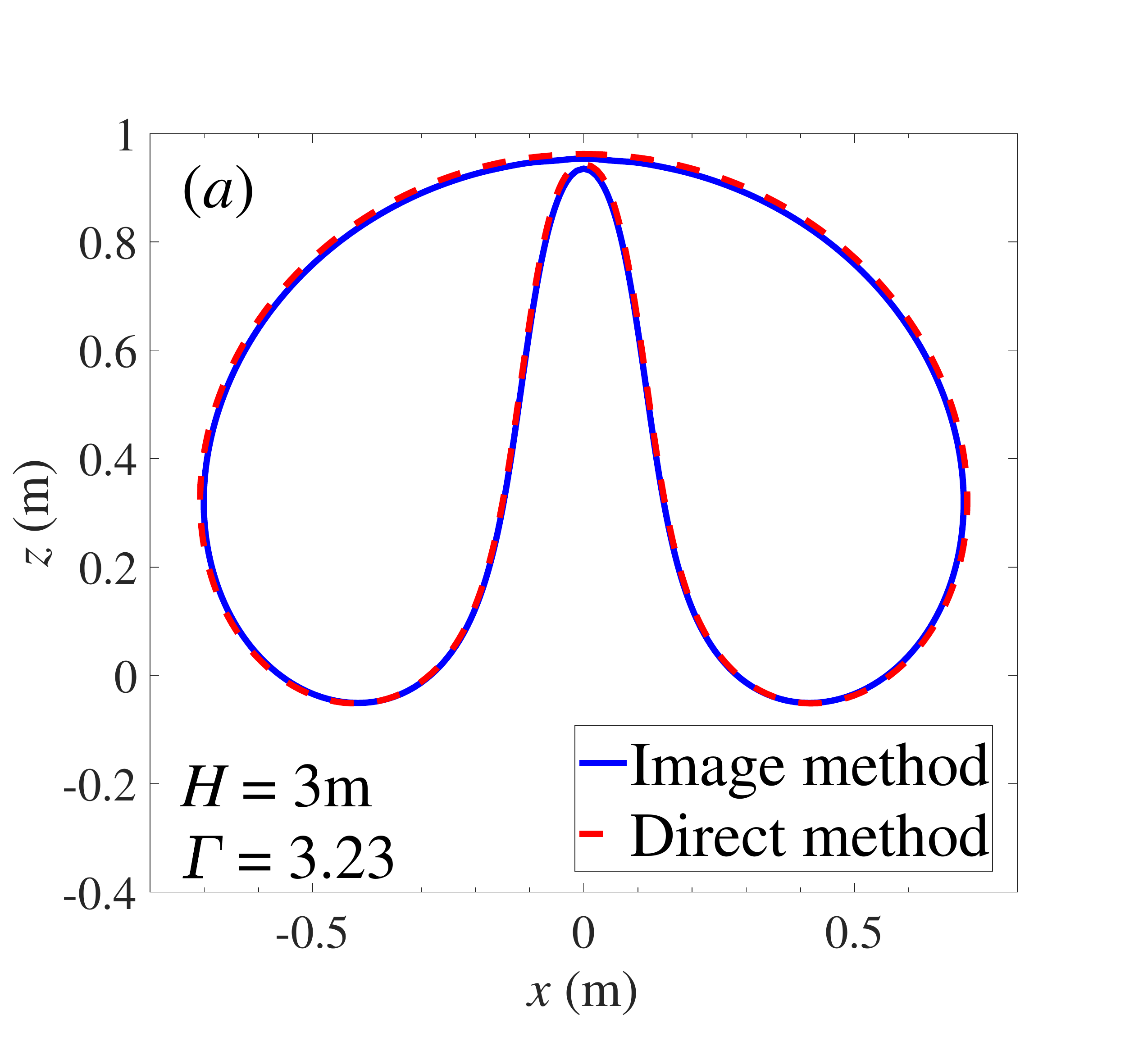}
	\includegraphics[width=6.5cm]{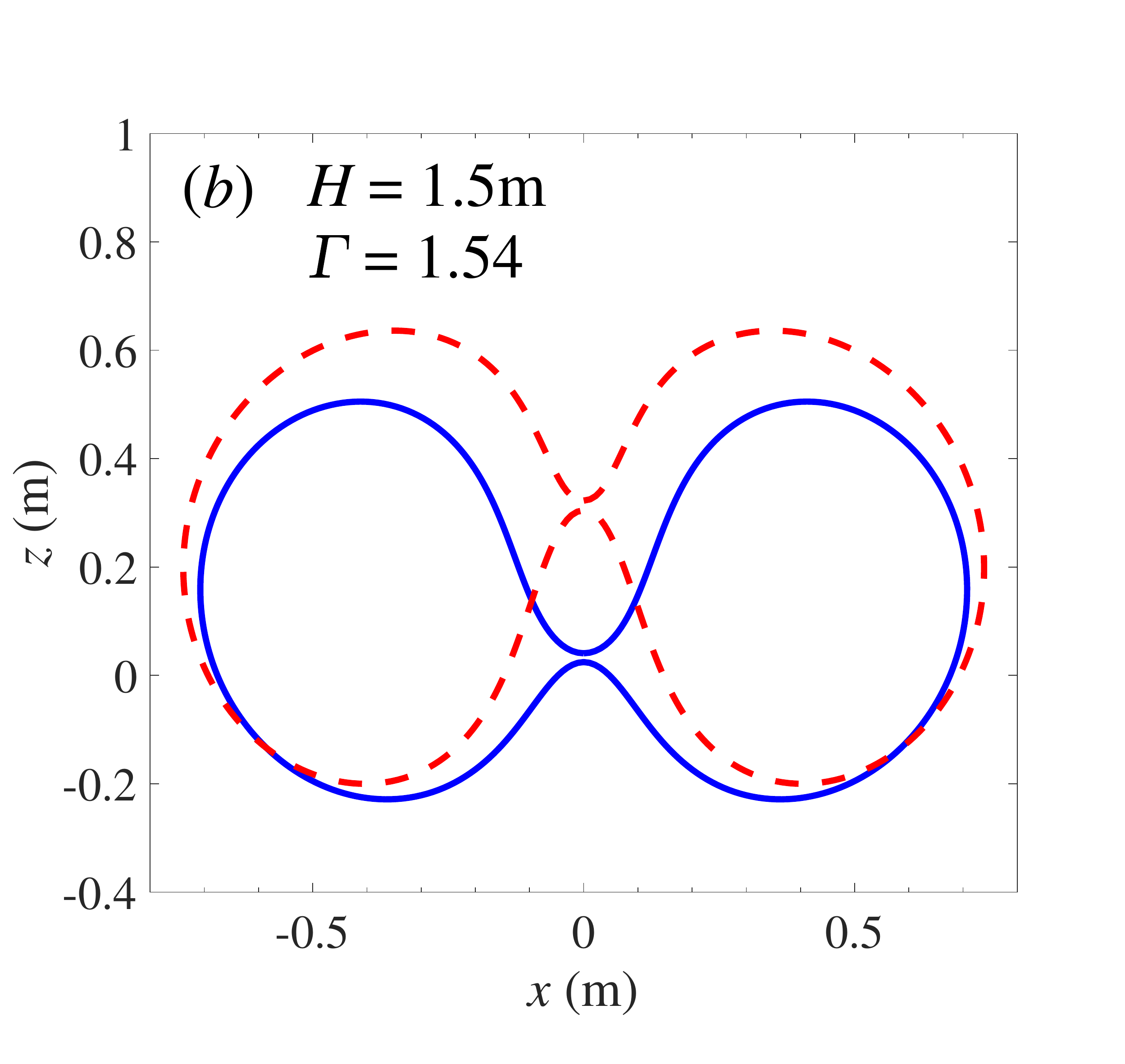}
	\caption{Comparison of bubble shapes at the moment just before jet penetration between the image method and the direct method for two different airgun firing depths and resulting the standoff parameter $\it\Gamma$ (defined as $\it\Gamma = H/R\rm_ m$), namely $H = 3{\rm m}, R_{\rm m} = 0.93{\rm m}$ and $\it\Gamma \rm= 3.23$ in (a) and $H = 1.5{\rm m},R_{\rm m} = 0.98{\rm m}$ and $\it\Gamma \rm = 1.54$ in (b). }\label{Fig:6}
\end{figure}

\subsection{Effects of the airgun firing depth}
\label{S:3-4}

\begin{figure}[htbp]
	\centering	
	\includegraphics[width=10cm]{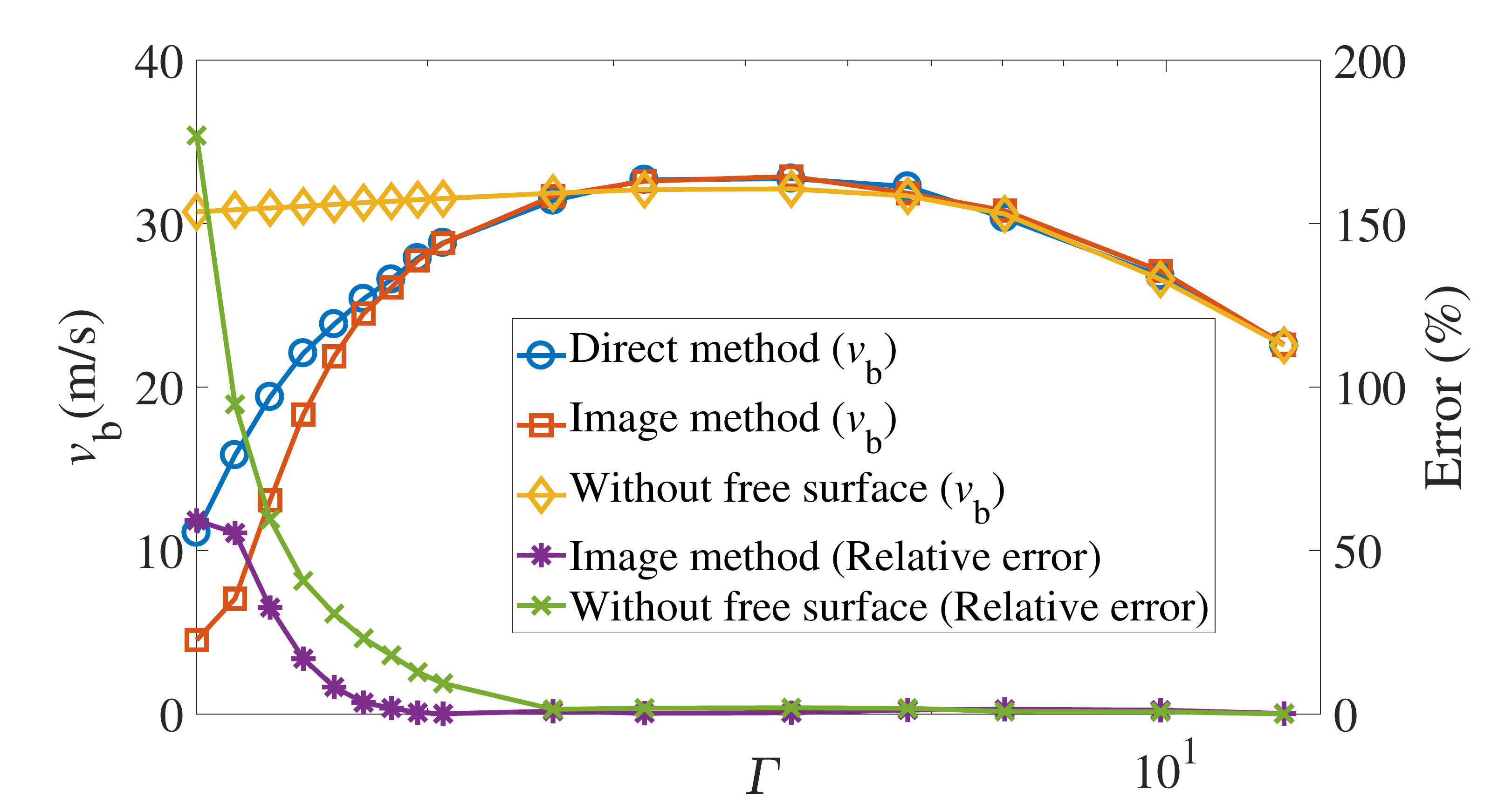}
	\caption{Comparison of the bubble jet velocity (defined as the bubble bottom velocity at the jet penetration moment) and accuracy between the image method and the direct method at different dimensionless airgun firing depth $\it\Gamma = H/R\rm_ m$.}\label{Fig:7}
\end{figure}

Many studies have been carried out for bubble-free-surface interactions \cite{Blake1987,Wang1996,Pearson2004}. Their main concern is the dynamic behaviors of a cavitation bubble and the free surface spike. However, the pressure wave generated by a gas bubble beneath a free surface has received little attention. There are two methods that incorporate the free surface effect into the numerical model. The first one is the ``image method'' \cite{LiGF2011,Cox,Klaseboer2005,ZhangS2018}, in which the free surface effect is modelled using a modified Green function $G(\textbf{\textit{r}},\textbf{\textit{q}})=1/|\textbf{\textit{r}}-\textbf{\textit{q}}|-1/|\textbf{\textit{r}}-\textbf{\textit{q}}^\prime|$. This method reduces computational cost but is only valid when the free surface remains relatively flat \cite{Klaseboer2005}. In the second  ``direct method'' \cite{Li-OE2018,Blake1987},  the free surface is modelled explicitly and the fully nonlinear boundary conditions are imposed. The simple Green function, $G(\textbf{\textit{r}},\textbf{\textit{q}})=1/|\textbf{\textit{r}}-\textbf{\textit{q}}|$, is used in the  ``direct method''. It is worth exploring the applicability of the image method.  First of all, the two methods are used to calculate the above ``standard case'' and the comparison of bubble profiles at the jet penetration moment is given in Figure \ref{Fig:6}(a). Very similar results are obtained from these two methods since the dimensionless standoff parameter
\begin{equation} 
\it\Gamma = H/R\rm_ m\label{Equation:20}
\end{equation}
 is 3.23, indicating a very weak bubble-free-surface interaction in this case. $\it\Gamma$ is a commonly used parameter in bubble dynamics because the bubble-free-surface interaction is highly dependent on it \cite{Zhang2015,Li-OE2018,Blake1987,Wang1996}. As the airgun firing depth ($H$) decreases, the bubble-free-surface interaction becomes stronger and the difference between these two methods is expected to increase, see the comparison for the $H \rm = 1.5\ m$ ($\it\Gamma \rm = 1.54$) case in Figure \ref{Fig:6}(b). Compared with the results obtained from the direct method, the jets obtained from the image method have broader widths and the impact point is lower. In such small $\it\Gamma$ cases, the direct method with higher accuracy should be used. To explore the scope of application of the image method, a series of simulations are carried out with a larger range of $\it\Gamma$. The relative error of the bubble jet velocity (the velocity of the bubble bottom at the jet penetration moment) at the jet penetration moment between these two methods is shown in Figure \ref{Fig:7}. The results obtained from a model that turns off the free surface effect are also given. We find that the relative error of the image method can be controlled within 5\% if $\it\Gamma \rm > 1.9$ and 1\% if $\it\Gamma \rm > 2.1$. Besides, the influence of a free surface on the bubble dynamics can be neglected if $\it\Gamma \rm > 6$. This finding provides a reference for the future modelling.

\begin{figure}[htbp]
	\centering	
	\includegraphics[width=6.5cm]{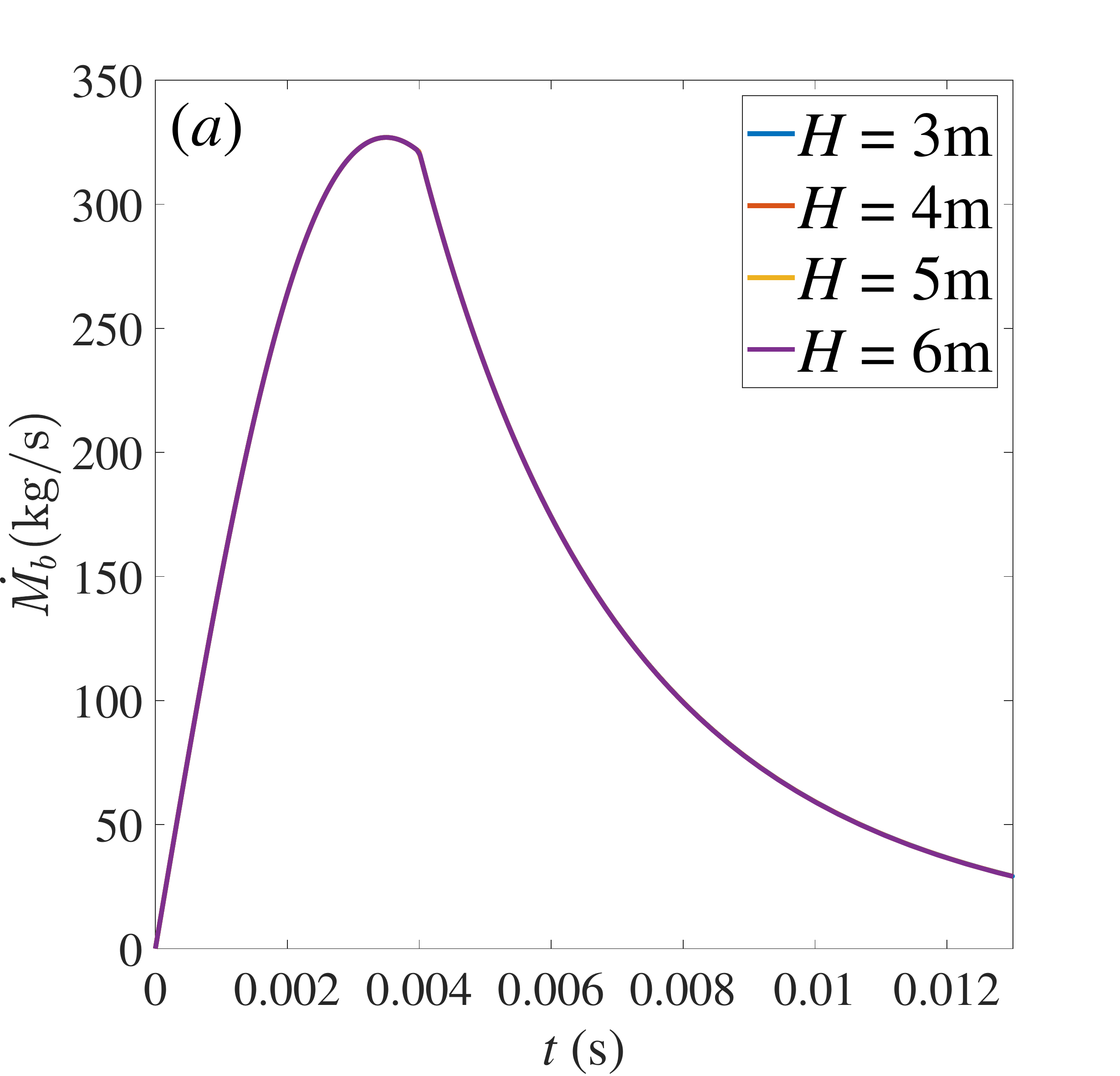}
	\includegraphics[width=6.5cm]{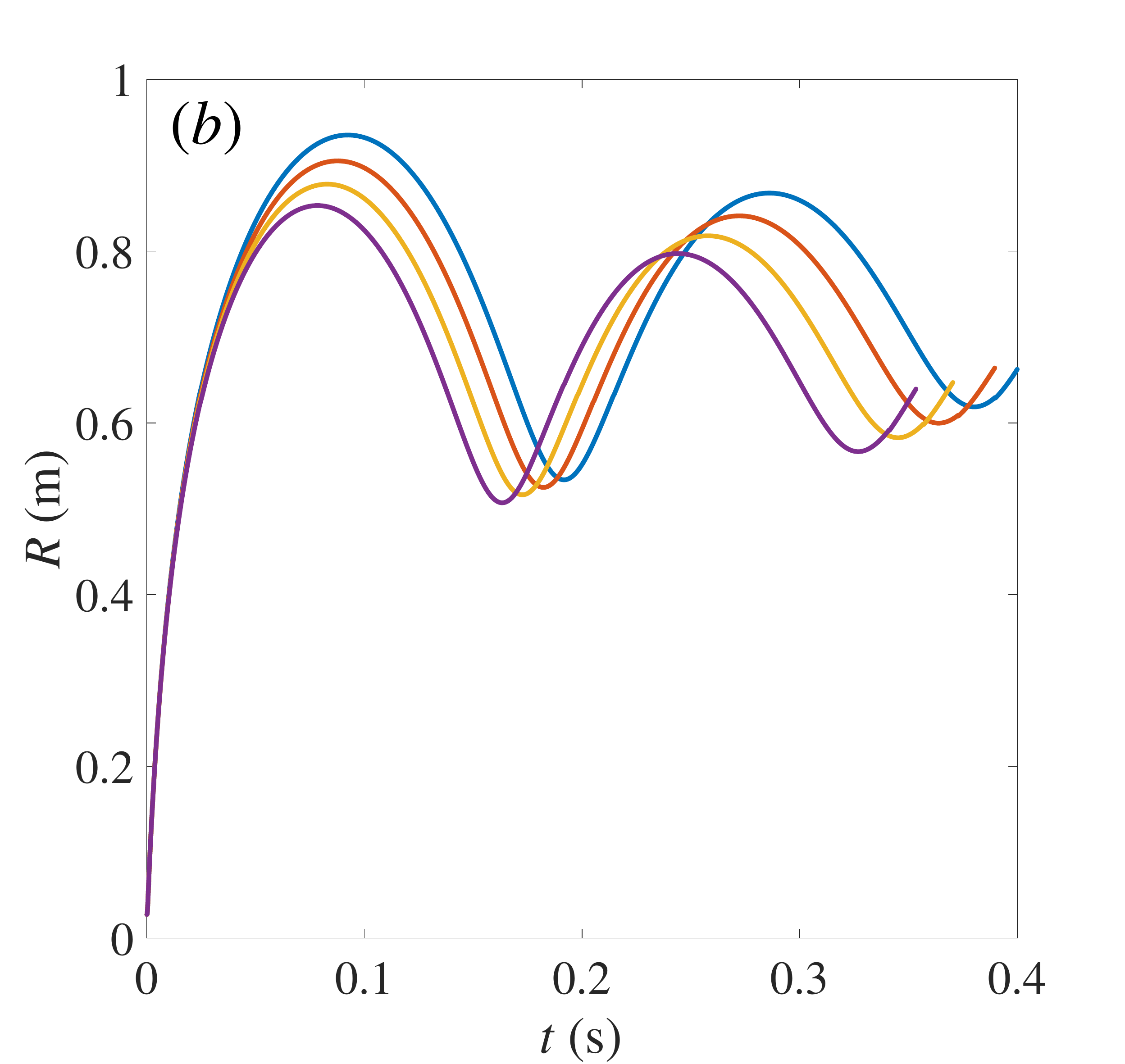}
	\includegraphics[width=6.5cm]{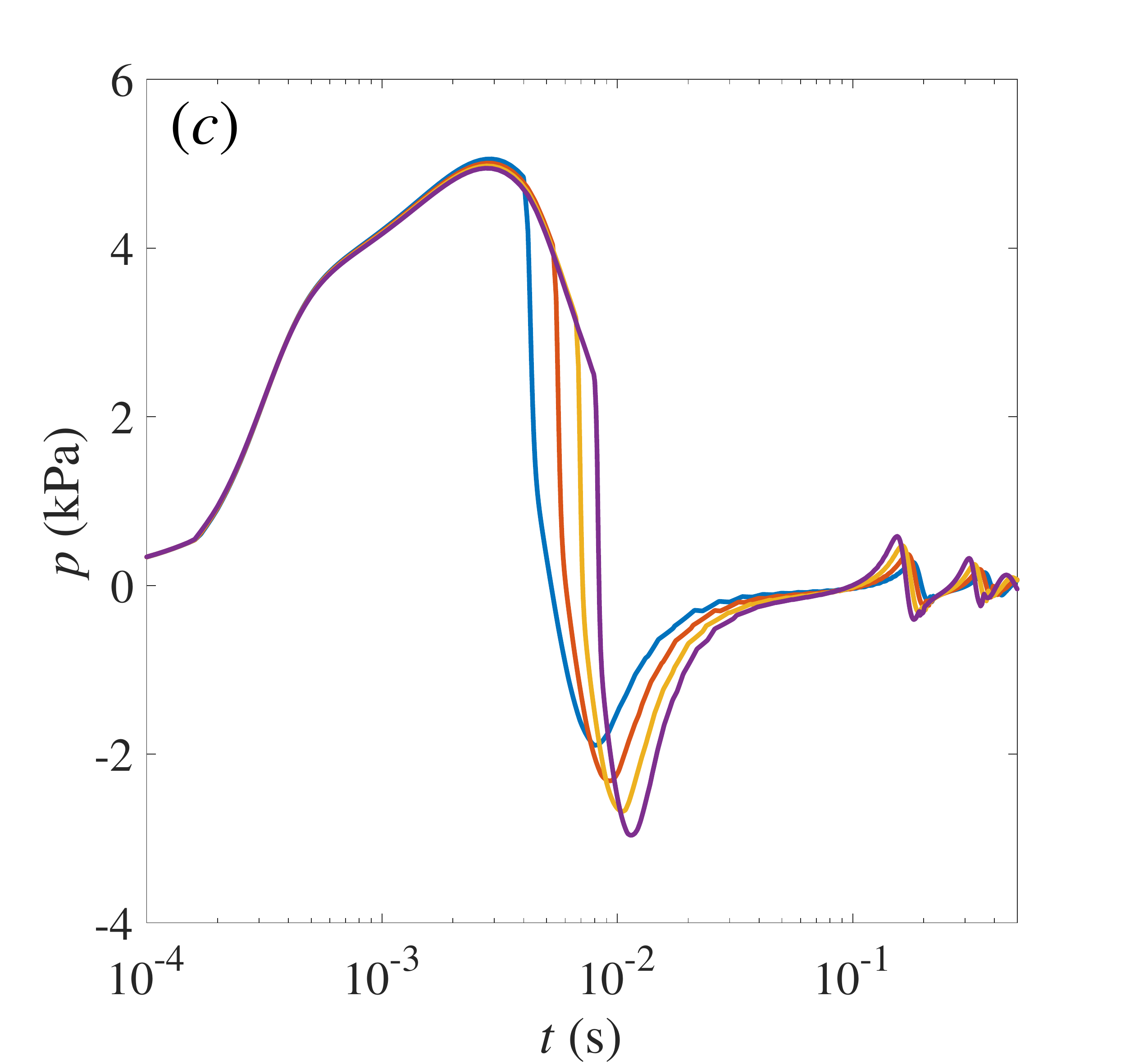}
	\includegraphics[width=6.5cm]{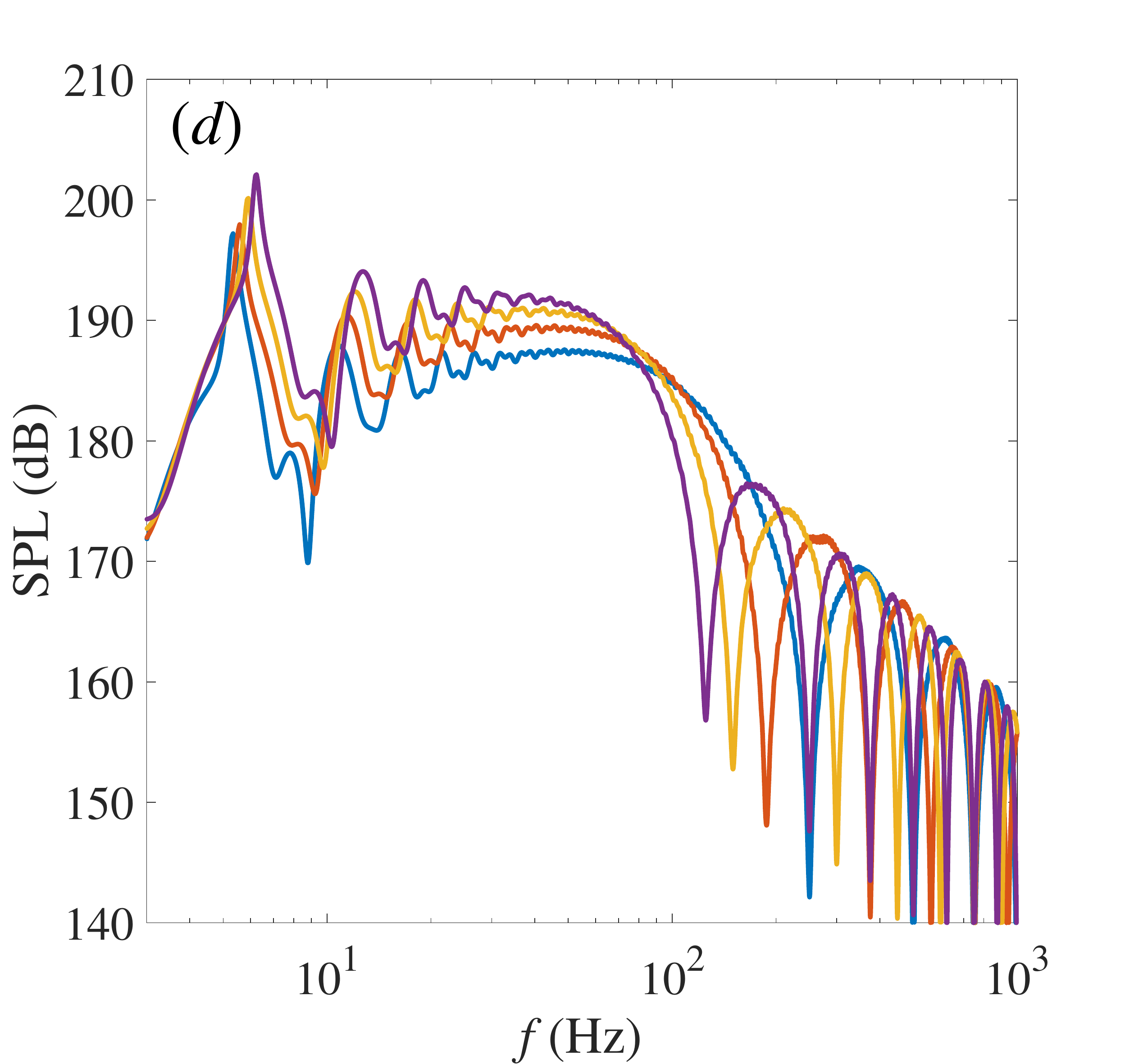}
	\caption{The effects of the airgun firing depth. (a) Mass flow rate from the airgun chamber to the bubble, (b) equivalent bubble radius, (c) far-field pressure wave (100 m below the initial airgun-bubble center), (d) amplitude spectrum. The parameters are kept the same as in the case of Figure \ref{Fig:3}(a).}\label{Fig:8}
\end{figure}

In the following, we present four different simulations which are conducted with the airgun firing depth varying from 3 to 6 m. Other parameters are the same as in the ``standard case''. The direct method is used here. Figure \ref{Fig:8}(a) shows the comparison of the air release rate between these four cases. The airgun firing depth in this range has little effect on the transient air release process. If we further increase the airgun firing depth to 30 m, the maximum value of $\dot{M\rm{_b}}$ only varies for 0.1\%. Although the bubble has nearly the same initialization phase, the bubble undergoes different oscillation processes, as shown in Figure \ref{Fig:8}(b). The hydrostatic pressure increases as the airgun go deeper, thus the bubble achieves a smaller maximum radius and shorter period. Figure \ref{Fig:8}(c) gives the far-field (100 m below the initial airgun-bubble center) pressure waves for different $H$. The pressure peak is reached at the same time in all cases as the air release phase is marginally affected by the airgun firing depth. However, the sudden drop of the pressure occurs earlier with a smaller airgun firing depth because the signal reflected from the free surface arrives at $\Delta t = 2H/c$ after the direct wave. Besides, the second pressure peak varies significantly with $H$ due to the superposition of the direct signal and the reflected signal. The SPL of the pressure waves in Figure \ref{Fig:8}(c) are given in Figure \ref{Fig:8}(d). In the low-frequency range, the amplitude of SPL increases with $H$ since the second pressure peak that mainly contributes to the low-frequency waves increases with $H$. However, the effective bandwidth (defined as the width of the frequency domain where the amplitude drops 6 dB below its maximum value) greatly decreases with $H$. The first notch is roughly located at the frequency of $f=1/\Delta t=c/2H$ \cite{ZhangS2018}. Therefore, the airgun firing depth primarily controls the effective frequency bandwidth and also affects the amplitudes of the low-frequency pressure waves.

\subsection{Effects of the heat transfer}
\label{S:3-5}

\begin{figure}[htbp]
	\centering	
	\includegraphics[width=6.5cm]{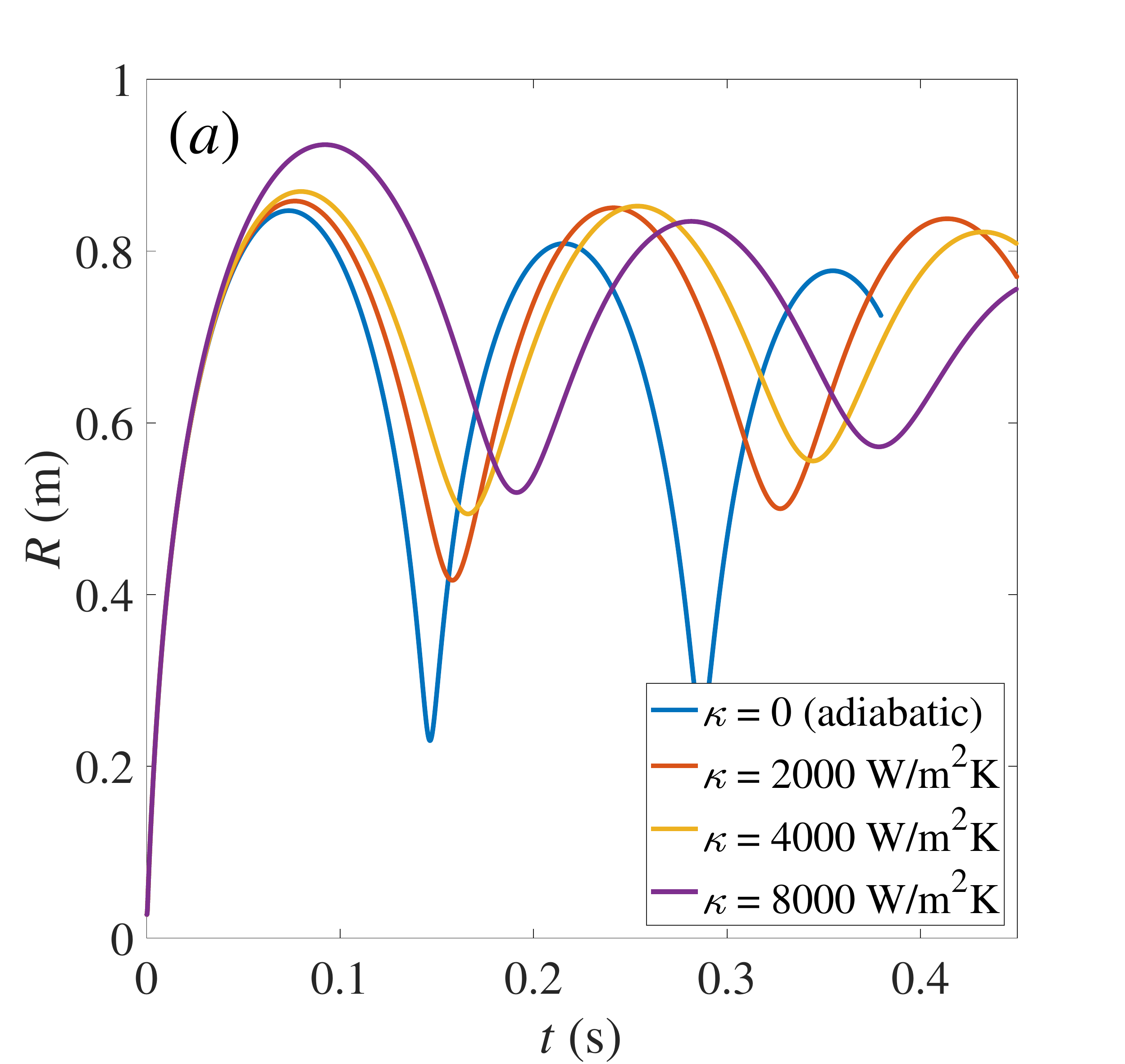}
	\includegraphics[width=6.5cm]{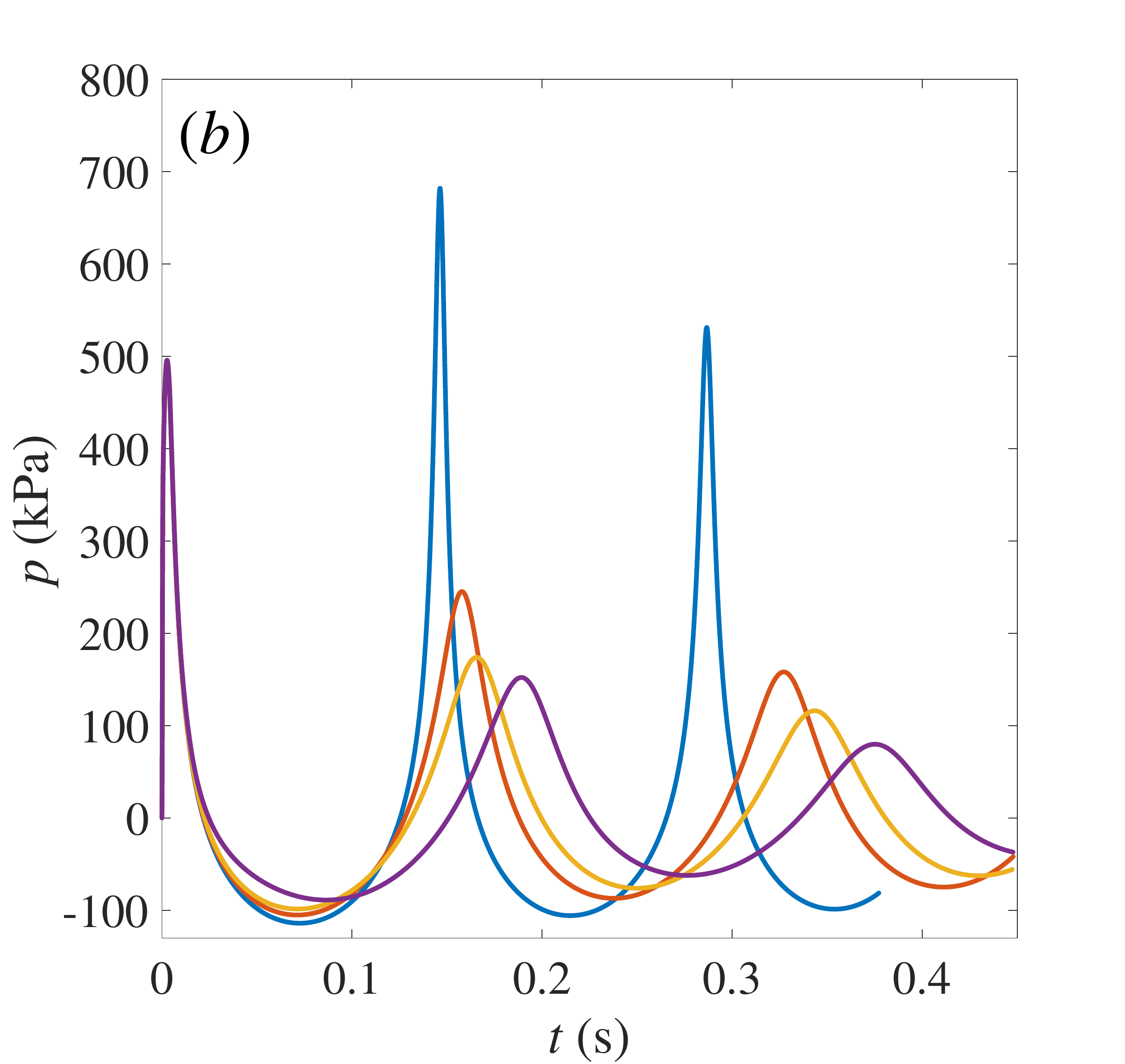}
	\includegraphics[width=6.5cm]{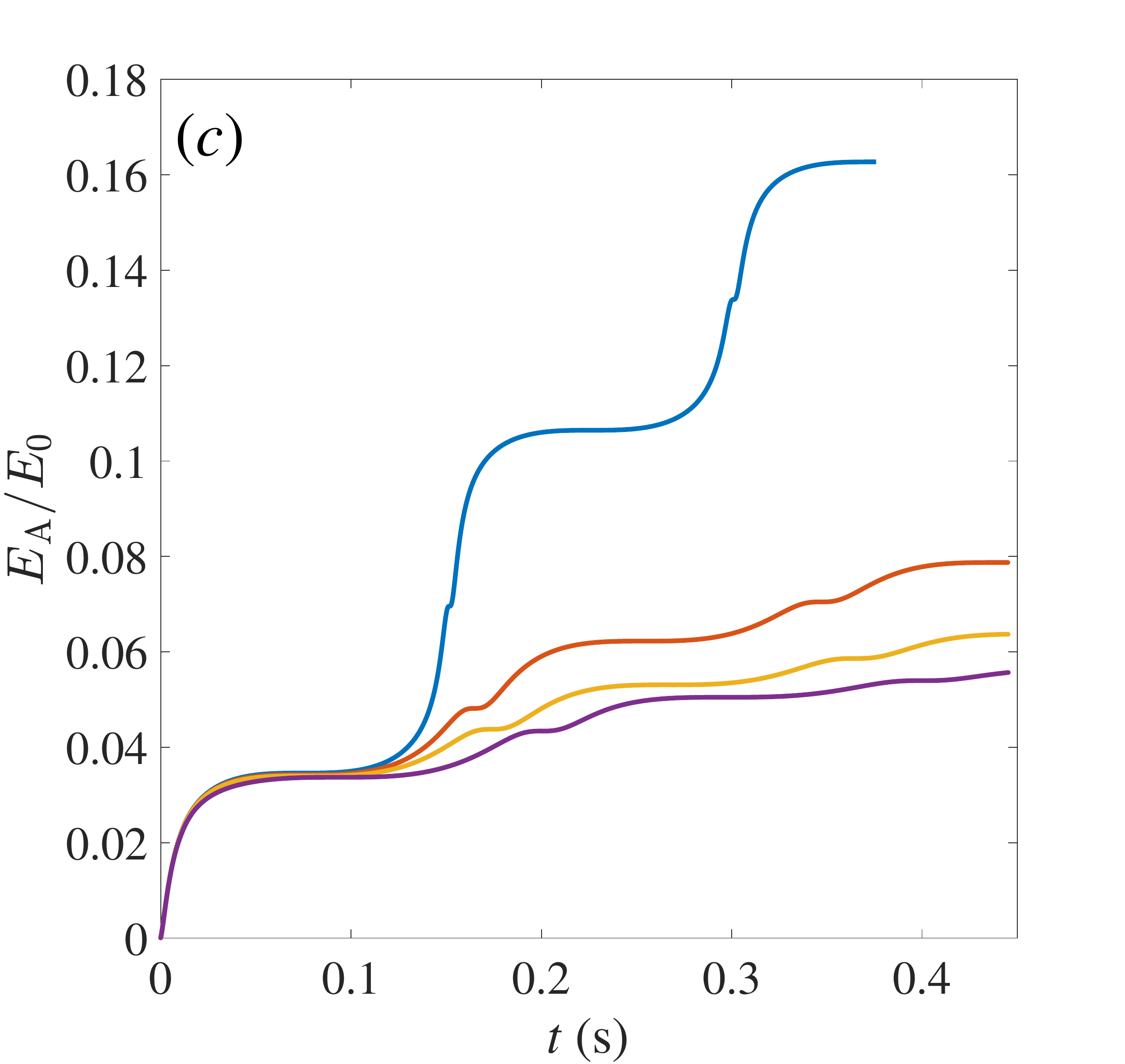}
	\caption{The effects of heat transfer. (a) Time evolution of the bubble radius, (b) time evolution of the near-field pressure wave (1 m away from the initial airgun-bubble center), (c) time evolution of the energy loss due to acoustic radiation. The parameters for the intial bubble are kept the same as in the case of Figure \ref{Fig:3}(a).}\label{Fig:9}
\end{figure}

Following the spherical bubble models in previous studies \cite{Laws1990,Graaf2014,ZhangS2017}, the heat transfer effect is also incorporated in our BI model. de Graaf \cite{Graaf2014} argued that the heat transfer is most likely the primary cause of bubble damping. For conventional airgun-bubbles, the thickness of the boundary layer varies during bubble oscillations but is in the order of 100$\mu m$ \cite{Laws1990,Graaf2014,Laws1991}, which is much smaller than the bubble size. The heat conducted across the boundary layer is assumed to be conducted away instantly into the bulk fluid. As previously done in theoretical models \cite{Laws1990,LiGF2011,ZhangS2017}, a simple model is used here to consider the heat transfer across the bubble-liquid interface, in which a constant  heat transfer coefficient $\kappa$ term in Equation \ref{Equation:16} is the only factor that controls the intensity of heat transfer. Based on the ``standard case'', we conduct four different cases with $\kappa$ being 0, 2000, 4000 and 8000 $\rm W/m^2K$, respectively. The bubble radius variations and the near-field pressure wave are given in Figure \ref{Fig:9}. We assume that the initial temperature of the air equals the ambient temperature, then the bubble cools down during the expansion and thus absorbs heat from the environment. Consequently, the bubble attains a larger maximum radius and longer oscillating period as $\kappa$ (heat transfer intensity) increases. The bubble collapsing is weakened as the heat is released from the bubble at the final collapse stage, thus the minimum bubble radius generally increases as the $\kappa$ increases. 

In the long term, the bubble oscillations decay faster as the heat transfer becomes stronger, as exhibited in both the bubble radius and the pressure wave dynamics. Figure \ref{Fig:9}(c) shows the time evolution of the energy loss due to acoustic radiation, which is calculated from Equation (\ref{Equation:acoustic}). $E{\rm_A}/E_0$ increases by approximately 3.3\% during the first expansion phase of the bubble. Thereafter, the increase of $E{\rm_A}/E_0$ is evident during very short periods in the vicinity of each minimum volume of the bubble. For the adiabatic case ($\kappa = 0$), the pressure wave emission is the only mechanism for energy decay. $E{\rm_A}/E_0$ increases to 10.6\% and 16.2\% at the ends of the first and second collapse phases, respectively. For the other three cases with heat transfer effects, the bubble collapse is weakened and the associated $E{\rm_A}/E_0$ decreases consequently. The maximum $E{\rm_A}/E_0$ is less than 8\% after the second oscillation cycle of the bubble. For the airgun type discussed in this study, the associated Mach number is much smaller than 1, thus the compressibility of the liquid does not seem to play a significant role in the bubble energy decay. This finding is consistent with previously published literature \cite{Ziolkowski1970,Graaf2014}. The compressibility of the liquid is the most important mechanism responsible for the energy decay of underwater explosion bubbles and cavitation bubbles \cite{WangQX2016JFM}.

\subsection{Effects of gravity}
\label{S:3-6}

Given the environmental damage that arises from the high-frequency components of the deep-sea seismic survey, it is desired to achieve a higher amplitude of low-frequency acoustic waves by using larger volume airguns. The gravity (buoyancy) effect is however expected to play a more important role for these larger bubbles. In this section, we will study the influence of gravity on the bubble dynamics and the pressure wave emission. To simplify the discussion, we turn off the effects of the  free surface, heat transfer, liquid compressibility and gas ejection in the model. For better comparison, all physical quantities are converted into dimensionless form using three fundamental quantities, namely, the maximum bubble radius $R\rm _m$, the hydrostatic pressure at the depth of the bubble inception point ($p{\rm _\infty} = p_{\rm atm}+\rho g H$, where $p_{\rm atm}$ is the atmospheric pressure) and the liquid density surrounding the bubble $\rho$. The dimensionless initial pressure and radius of the bubble are chosen as 20 and 0.248, respectively. The time is scaled by $R_{\rm m} \sqrt{\rho / p{\rm _\infty}}$. All the discussions in this section are in dimensionless quantities. The asterisk * denotes a dimensionless quantity. With gravitational effect included the dynamic boundary condition in the dimensionless form is given by:
\begin{equation} 
\frac{{\rm d}{\varphi^\ast}}{{\rm d}t^\ast}=\frac{1}{2}{|{\nabla \varphi^\ast}|^2}+1-p_{0}^\ast\left(\frac{V_0^\ast}{V^\ast} \right) ^\gamma-\frac{1}{Fr^2}z^\ast.\label{Equation:24}
\end{equation}

With the simplification of the problem, the bubble dynamics are now uniquely determined by the Froude number
\begin{equation} 
Fr = \sqrt{p_\infty/{\rho gR_m}}.\label{Equation:25}
\end{equation}

\begin{figure}[htbp]
	\centering\includegraphics[width=13cm]{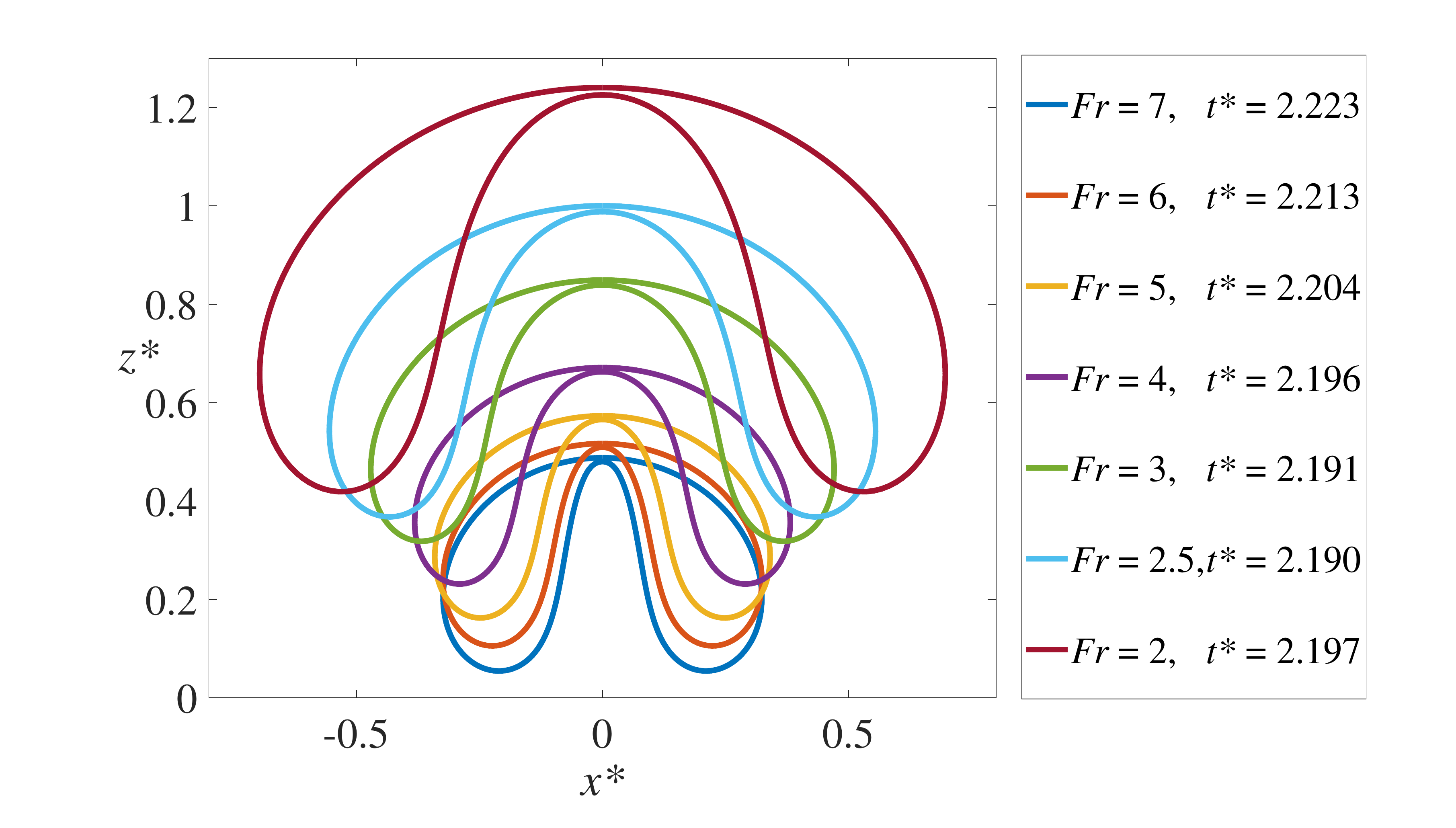}
	\caption{Bubble shapes at the jet penetration time for different Froude numbers.}\label{Fig:10}
\end{figure}

To reveal the dependence of the bubble dynamics on $Fr$, a parametric study is carried out in the regime $2 \leq Fr \leq 7$.  Figure \ref{Fig:10} shows the comparison of bubble profiles at the jet penetration moment. The bubble migration increases as $Fr$ decreases, indicating a stronger gravity/buoyancy effect. It is also noted that the bubble volume at this moment decreases as $Fr$ increases. Generally, jet penetration occurs earlier as $Fr$ decreases. The most important feature is that the jet is more vigorous for decreasing $Fr$, which alters the subsequent bubble collapse behavior and the pressure wave emission. Compared with the $Fr = 2.5$ case, the jet penetration in the $Fr = 2$ case delays a little. This is because the decreasing ambient pressure during bubble migration slows down the bubble collapse.

\begin{figure}[htbp]
	\centering	
	\includegraphics[width=13cm]{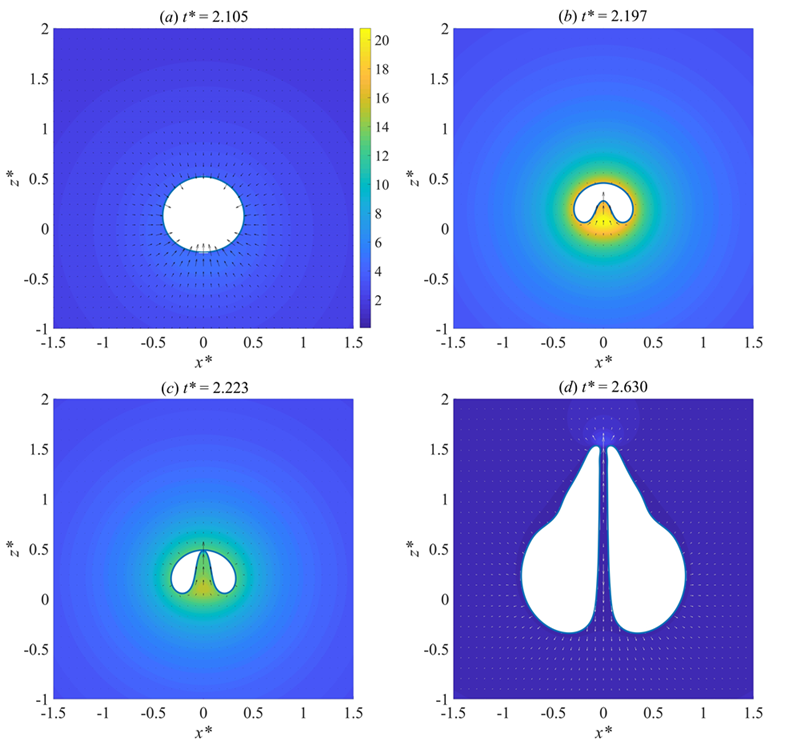}
	\caption{The pressure contours and velocity fields at the final collapse and early rebound stages of the bubble with $Fr = 7$. The bubble maintains a spherical spherical shape during most of the first cycle and the upward jet forms at the end of the first cycle and penetrates the upper surface during the rebound phase. }\label{Fig:11}
\end{figure}

In the following, we will analyze the detailed bubble motion for two extreme cases, i.e., the $Fr = 7$ and $Fr = 2$ cases. Figure \ref{Fig:11} shows the pressure contours and velocity fields surrounding the bubble for the $Fr = 7$ case. The bubble suffers a relatively weak gravity/buoyancy effect, thus the bubble keeps a spherical shape during most of the first cycle. The bubble obtains its minimum volume before jet penetration, as shown in Figure \ref{Fig:11}(b). Except for a localized high pressure region at the bottom of the bubble, the pressure is spherically symmetric near the bubble surface. Therefore, the spherical bubble theory is an appropriate way to model small-scale airgun-bubbles. The jet impact also causes a localized high pressure region around the jet tip, however, this does not seem to play a role in the far field pressure.

\begin{figure}[htbp]
	\centering	
	\includegraphics[width=13cm]{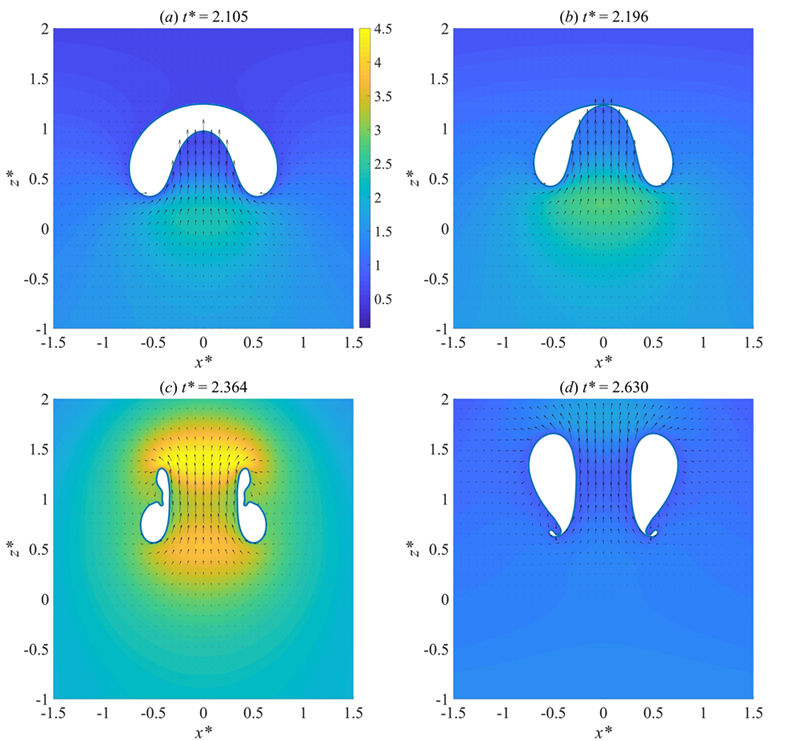}
	\caption{The pressure contours and velocity fields at the final collapse and early rebound stages of a bubble with $Fr = 2$. Subject to stronger buoyancy effect, the liquid jet forms relatively earlier than that in the $Fr = 7$ case. When the bubble becomes toroidal, there exists an annular neck on the bubble surface, which is propagating downward and finally leads to the splitting of the bubble.}\label{Fig:12}
\end{figure}

Figure \ref{Fig:12} shows the pressure contours and velocity fields of the $Fr = 2$ case. The bubble jet forms earlier than that in the $Fr = 7$ case, as shown in Figure \ref{Fig:12}(a). The focusing flow below the bubble causes a localized high-pressure region, which drives the jet upward continuously. After the jet penetration, the bubble keeps collapsing and reaches its minimum volume at $t^\ast = 2.364$. Two high-pressure regions can be observed around the bubble top and bottom. There exists an annular neck on the toroidal bubble surface, which is propagating downward and finally leads to the splitting of the bubble, as shown in Figure \ref{Fig:12}(d). 

By comparing the above two extreme cases, it is clear that the bubble dynamic is highly dependent on $Fr$. More specifically, the bubble becomes less spherical (reflected in a stronger jet) during the collapse phase with a smaller $Fr$. The kinetic energy associated with the jet formation increases from 0.43 to 2.05 as $Fr$ decreases from 7 to 2. Consequently, the minimum bubble volume increases as $Fr$ decreases. Therefore, the maximum pressure induced by the bubble at the minimum bubble volume increases with $Fr$. 

\begin{figure}[htbp]
	\centering\includegraphics[width=10cm]{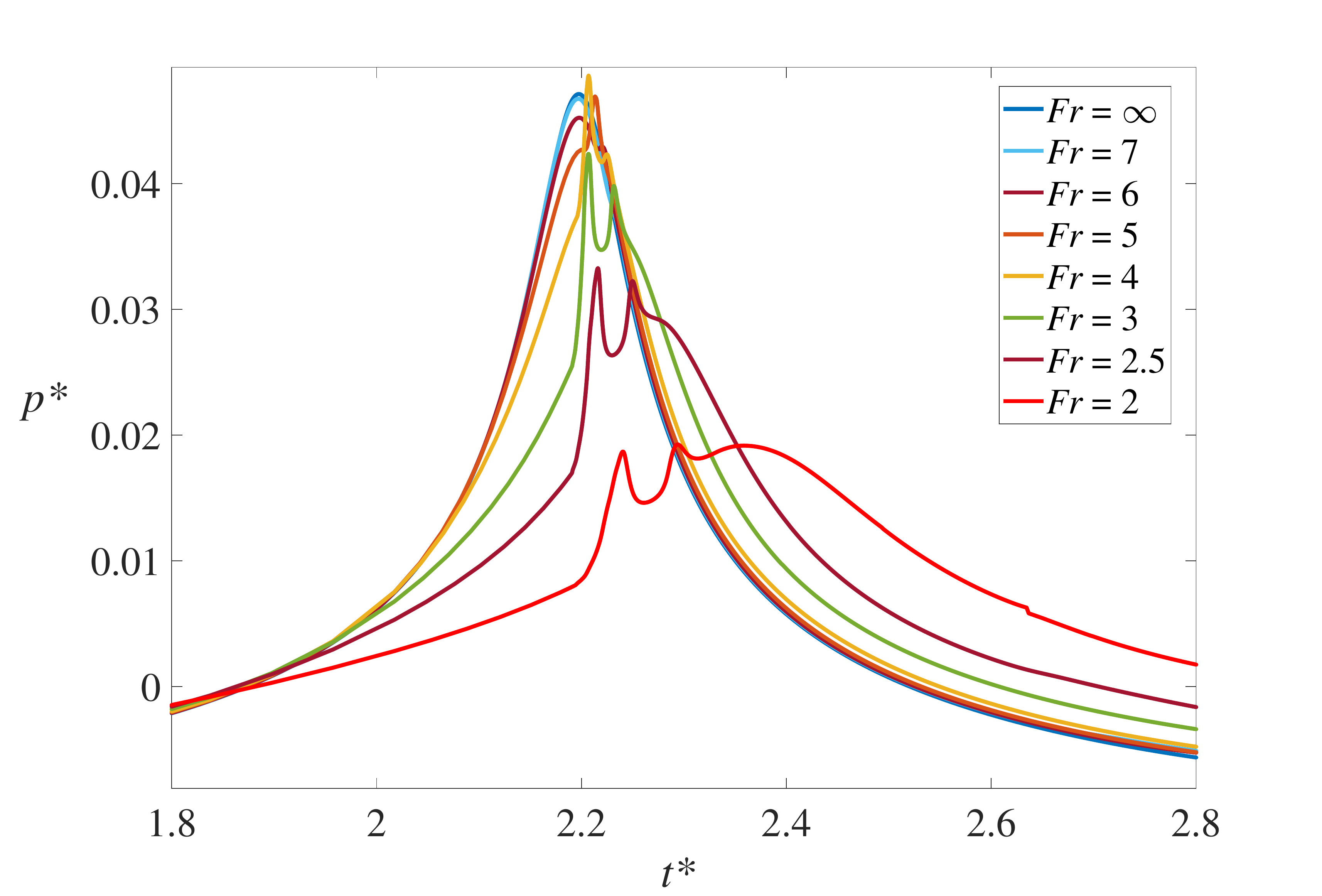}
	\caption{Comparison of the pressure wave in the far-field ($D^* = 100$) generated by bubbles with different $Fr$. }\label{Fig:13}
\end{figure}

Figure \ref{Fig:13} shows the pressure waves in the far-field generated by bubbles with different $Fr$. The $Fr=\infty$ case represents the spherical bubble situation. The waveforms of $Fr = 7$ and $Fr = \infty$ cases are almost identical, indicating that the gravity plays a minor role when $Fr \ge 7$ and the spherical bubble theory is an appropriate description. This is why spherical bubble theories could be successfully applied in small scale airgun-bubble modelling during the past half-century. The gravity effects obviously increase with decreasing $Fr$ and the waveform deviates from the $Fr = \infty$ case gradually. The pressure peak decreases with $Fr$ and multiple pressure peaks can be observed when $Fr \leq 4$. Generally, the first pressure peak relates to the violent jet impact and the subsequent pressure peak relates to the combination of the high-pressure gas and the decaying jet flow \cite{Li_oe2016}. Therefore, for a large scale airgun-bubble, describing the bubble oscillation with the spherical bubble theory will neglect the significant jetting behavior during the collapse phase, which has a great effect on the pressure wave emissions.

It is worth mentioning that all the discussion in this section is within the dimensionless framework as given above. Note that the bubble size is the most important quantity that alters $Fr$, namely $Fr$ decreases as $R_{\rm m}$ increases. If we consider the same dimensional distance, the maximum pressure induced by the bubble increases with $R_{\rm m}$.

\section{Conclusions and outlook}
\label{S:4}

In this study, a non-spherical airgun-bubble model has been established based on the boundary integral method in conjunction with an improved air release model. The effects of gravity, free ocean surface, liquid compressibility, heat transfer, air release and port opening process are considered in the numerical model, which incorporates more physical details than previously done. The model is validated by comparisons with three experiments. The highly non-spherical bubble behavior of an electric discharge bubble is well captured by the BI code, including the bubble jetting, toroidal bubble splitting and the interaction between two sub-toroidal bubbles. Also, the pressure wave emissions generated by real airgun-bubbles are reproduced by the numerical model, including the pressure peaks and the bubble period. Thereafter, parametric studies are carried out to reveal the dependence of airgun-bubble dynamics on governing factors. The main findings are given as follows.

\begin{enumerate}[\indent(1)]

\item The accuracy of the ``image method'' that considers the free surface effect by using a negative image of the bubble is compared with the ``direct method''. We found that the relative error of the image method can be controlled within 5\% if $\it\Gamma \rm> 1.9$ and 1\% if $\it\Gamma \rm > 2.1$ ($\it\Gamma = H/R\rm_ m$, where $H$ is the airgun firing depth and $R_{\rm m}$ is the maximum bubble radius). The influence of a free surface on the bubble dynamics can be neglected if $\it\Gamma \rm > 6$.

\item The airgun firing depth $H$ in the conventional range of airgun use (less than 30 m) has little effect on the transient air release process, thus the first pressure peak is not much affected by $H$. The second pressure peak increases significantly with $H$ due to the superposition of the direct signal and the reflected signal, which results in a higher amplitude of low-frequency pressure waves. However, the effective frequency bandwidth decreases with increasing $H$.

\item  A smaller port opening time $T_{\rm open}$ leads to a more violent air release process and a higher first pressure peak, which primarily contributes more to high-frequency pressure waves but has little influence on the low-frequency pressure waves. This finding provides a reference for the future design of environmentally friendly airguns.

\item The non-spherical bubble dynamics is highly dependent on $Fr$. As $Fr$ decreases, the jet contains higher kinetic energy, thus the bubble occupies lower potential energy and leads to weaker energy focusing and the maximum pressure induced by the collapsing bubble decreases. Gravity plays a minor role when $Fr \ge 7$ and the spherical bubble theory becomes an appropriate description.

\item Our BI simulations suggest that bubble splitting is likely to occur during the toroidal bubble stage, which will constitute to energy dissipation in airgun-bubble dynamics. 
\end{enumerate}

Considering the airgun size is much smaller than the bubble, the effect of the airgun body is neglected in the present study, which is a possible reason for the slight mismatch between the experiment and the simulation. Finally, as an outlook, we mention that the present BI code is suitable for simulating the nonlinear interaction between multiple bubbles, which is a relevant question in deep sea seismic survey.

\section{Acknowledgements}
\label{S:5}

The authors gratefully acknowledge SHELL for funding this work.





\bibliographystyle{model1-num-names}
\bibliography{sample}







\end{document}